\documentclass{aa}
\usepackage{graphicx}
\usepackage[varg]{txfonts}
\usepackage{longtable}
\usepackage{lscape}
\usepackage{natbib}
\usepackage{url}
\usepackage{color}
\usepackage{multirow,bigdelim}
\usepackage{cases}
\usepackage[colorinlistoftodos]{todonotes}
\bibpunct{(}{)}{;}{a}{}{,} 
\newcommand\kms{{\rm\,km\,s^{-1}}}

\newcommand\teff{T_{\rm eff}}

\usepackage{hyperref}
\hypersetup{
     breaklinks,
     colorlinks   = true,
     citecolor    = blue,
     urlcolor     = blue,
     linkcolor    = blue
}

\begin{document}

\title{
Chemical evolution of the Galactic bulge as traced by\\
microlensed dwarf and subgiant stars\thanks{Based on data obtained with the European Southern Observatory telescopes (Proposal ID:s 87.B-0600, 88.B-0349, 89.B-0047, 90.B-0204, 91.B-0289, 92.B-0626, 93.B-0700), the Magellan Clay telescope at the Las Campanas observatory, and the Keck~I telescope at the W.M. Keck Observatory, which is operated as a scientific partnership among the California Institute of Technology, the University of California and the National Aeronautics and Space Administration. }
\fnmsep
\thanks{Table~2 is available in electronic form at the CDS via anonymous ftp to \url{cdsarc.u-strasbg.fr (130.79.128.5)} or via \url{http://cdsweb.u-strasbg.fr/cgi-bin/qcat?J/A+A/XXX/AXX}.
}}
\subtitle{
VIII. Carbon and oxygen
}
\titlerunning{Chemical evolution of the Galactic bulge as traced by microlensed dwarf and subgiant stars. VIII.}

\author{
T.~Bensby\inst{1}
\and
A.~Gould\inst{2,3}
\and
M.~Asplund
\and
S.~Feltzing\inst{1}
\and
J.~Mel\'endez\inst{4}
\and
J.A.~Johnson\inst{2}
\and\\
S.~Lucatello\inst{5}
\and
A.~Udalski\inst{6}
\and
J.C.~Yee\inst{7}
 }

\institute{Lund Observatory, Department of Astronomy and 
Theoretical physics, 
Box 43, SE-221\,00 Lund, Sweden
\and
Department of Astronomy, Ohio State University, 140 W. 18th Avenue, 
Columbus, OH 43210, USA
\and
Max Planck Institute for Astronomy, K\"onigstuhl 17, D-69117 Heidelberg, Germany
\and
Departamento de Astronomia do IAG/USP, Universidade de S\~ao Paulo,
Rua do Mat\~ao 1226, S\~ao Paulo, 05508-900, SP, Brasil
\and
INAF-Astronomical Observatory of Padova, Vicolo dell'Osservatorio 5, 
35122 Padova, Italy
\and
Astronomical Observatory, University of Warsaw, Al. Ujazdowskie 4, 00-478 Warszawa, Poland
\and
Center for Astrophysics $|$ Harvard \& Smithsonian, 60 Garden St., Cambridge, MA 02138, USA
}


\date{Received 18 June 2021 / Accepted 07 September 2021}

 \abstract{
 Next to H and He, carbon is together with oxygen the most abundant elements in the Universe and are widely used when modelling the formation and evolution of galaxies and their stellar populations. For the Milky Way bulge, there are currently essentially no measurements of carbon in un-evolved stars, hampering our abilities to properly compare Galactic chemical evolution models to observational data for this still enigmatic stellar population.
 }
 {
 We aim to determine carbon abundances for our sample of 91 microlensed bulge dwarf and subgiant stars. Together with new determinations for oxygen this forms the first statistically significant sample of bulge stars that have C and O abundances measured, and for which the C abundances have not been altered by the nuclear burning processes internal to the stars.
 }
 {
 The analysis is based on high-resolution spectra for a sample of 91 dwarf and subgiant stars that were obtained during microlensing events when the brightnesses of the stars were highly magnified. Carbon abundances were determined through spectral line synthesis of six \ion{C}{i} lines around 9100\,{\AA}, and oxygen abundances using the three \ion{O}{i} lines at about 7770\,{\AA}. One-dimensional (1D) MARCS model stellar atmospheres calculated under the assumption of local thermodynamic equilibrium (LTE) were used and non-LTE corrections were applied when calculating the synthetic spectra for both C and O.
 }
 {
Carbon abundances was possible to determine for 70 of the 91 stars in the sample and oxygen abundances for 88 of the 91 stars in the sample. The [C/Fe] ratio evolves essentially in lockstep with [Fe/H], centred around solar values at all [Fe/H]. The [O/Fe]-[Fe/H] trend has an appearance very similar to that observed for other $\alpha$-elements in the bulge, with the exception of a continued decrease in [O/Fe] at super-solar [Fe/H], where other $\alpha$-elements tend to level out.
When dividing the bulge sample into two sub-groups, one younger than 8\,Gyr and one older than 8\,Gyr, the stars in the two groups follow exactly the elemental abundance trends defined by the solar neighbourhood thin and thick disks, respectively.
Comparisons with recent models of Galactic chemical evolution in the [C/O]-[O/H] plane shows that the models that best match the data are the ones that have been calculated with the Galactic thin and thick disks in mind. 
 }
 {
We conclude that carbon, oxygen, and the combination of the two supports the idea that the majority of the stars in the Galactic bulge has a secular origin, that is, being formed from disk material. We cannot exclude that there is a fraction of stars in the bulge that could be classified as a classical bulge population, but it has to be small. More dedicated and advanced models of the inner region of the Milky Way are needed to make more detailed comparisons to the observations. 
 }
   \keywords{
   Gravitational lensing: micro ---
   Galaxy: bulge ---
   Galaxy: formation ---
   Galaxy: evolution ---
   Stars: abundances
   }
   \maketitle

\section{Introduction}

Carbon and oxygen are, next to H and He, the most abundant elements in the Universe and knowing their origins and evolutions through cosmic time is important for many fields of astrophysics \citep[e.g.][]{trimble1975,chiappini2003,pavlenko2019}. For instance, when probing the chemical evolution of galaxies and their stellar populations the C/O abundance ratio is especially important as C and O are produced by different progenitors that operate on different time scales  \citep[e.g.][]{tinsley1979}. Oxygen is believed to be solely made in massive stars \citep[e.g.][]{talbot1974} and therefore the enrichment of oxygen to the interstellar medium occurs on short time scales. The origin of carbon is still debated, and it is not clear whether low-, intermediate-, or high-mass stars are the main sources of carbon enrichment \citep[e.g.][]{franchini2020}.

Carbon can, together with nitrogen, be used to probe the internal structure and evolution of stars as they are active ingredients in nuclear burning processes, and once stars reach the red giant phase the processed materials are dredged to the surface, altering the abundances of C and N \citep[e.g.][]{charbonnel1994,ryde2009,lagarde2019}. While this is an exciting probe of the internal structure of the stars, the drawback is that if one wants to probe the carbon abundance of the gas cloud that a star was formed from, giant stars are not reliable indicators of C and N, and instead one has to study less evolved, and less luminous, dwarf and subgiant stars. This means that studies of carbon in un-evolved stars have been limited to regions that are relatively close to the Sun \citep[e.g.][]{gustafsson1999,bensby2004,bensby2006,nissen2014,amarsi2019b,stonkute2020,franchini2020}, while there is a severe lack  of reliable carbon determinations in distant regions of the Milky Way, such as the Galactic bulge, to study Galactic chemical evolution. Currently only three un-evolved stars in the bulge have carbon measurements \citep{johnson2008,cohen2009} making it impossible to make comparisons to Galactic chemical evolution models in the [C/O]-[O/H] plane \citep{romano2020}. A statistically significant sample of stars in the bulge with proper C and O abundances is crucial for us to make progress in our understanding of the formation and evolution of the Galactic bulge and to determine whether it is a distinct stellar population of the Milky Way or a region of the Milky Way formed through secular evolution of the Galactic disk, as so called pseudo-bulge \citep[e.g.][]{kormendy2004}. Oxygen in itself is also an important, and common, diagnostic that has been used to identify and distinguish different stellar populations of the Milky Way \citep[e.g.][]{bensby2004} and constraining the star formation rate and the initial mass function \citep[e.g.][]{romano2005}. In addition,  from the variance of [O/Fe] in the Galactic disk, based on APOGEE spectra, \cite{bertrandelis2016} argue that the spread in [O/Fe] and other abundance ratios provide strong constraints on chemical evolution models.

In the years 2009-2015 we conducted a campaign to observe un-evolved dwarf and subgiant stars in the bulge at high spectral resolution while they were being microlensed. Our main objective was to determine ages for individual stars from isochrone fitting, and also to determine precise elemental abundances. In a series of papers we have done that for in total of 91 microlensed dwarf and subgiant stars \citep{bensby2009,bensby2010,bensby2013,bensby2017}. Most recently we presented what was at the time, and still remains, the most detailed study of Li in the Galactic bulge \citep{bensby2020}. In the current study we aim to reconcile the situation of the carbon abundance trend in the Galactic bulge and how it compares to carbon trends in the other Galactic stellar populations, in particular the thin and thick disks. At the same time we will determine oxygen abundances for all stars, and use these to create the important diagnostic $\rm [C/O]-[O/H]$ abundance trend plots that are vital when comparing models of Galactic chemical evolution to observations. This diagnostic plot is important because C and O yields do not depend on the uncertain location of the mass-cut in core-collapse SNe (at variance with Fe).

Section~\ref{sec:sample} summarises the main characteristics of the stellar sample and describes how it was obtained and analysed in previous papers; Sect.~\ref{sec:analysis} describes how the carbon and oxygen abundances were determined; Sect.~\ref{sec:results} presents the results for oxygen, carbon, and the carbon-over-oxygen ratio; Sect.~\ref{sec:discussion} makes a comparison to Galactic chemical evolution models; and Sect.~\ref{sec:summary} concludes with a summary of the results and our conclusions. 

\section{Stellar sample and stellar parameters}
\label{sec:sample}

\begin{figure}
\centering
\resizebox{\hsize}{!}{
\includegraphics{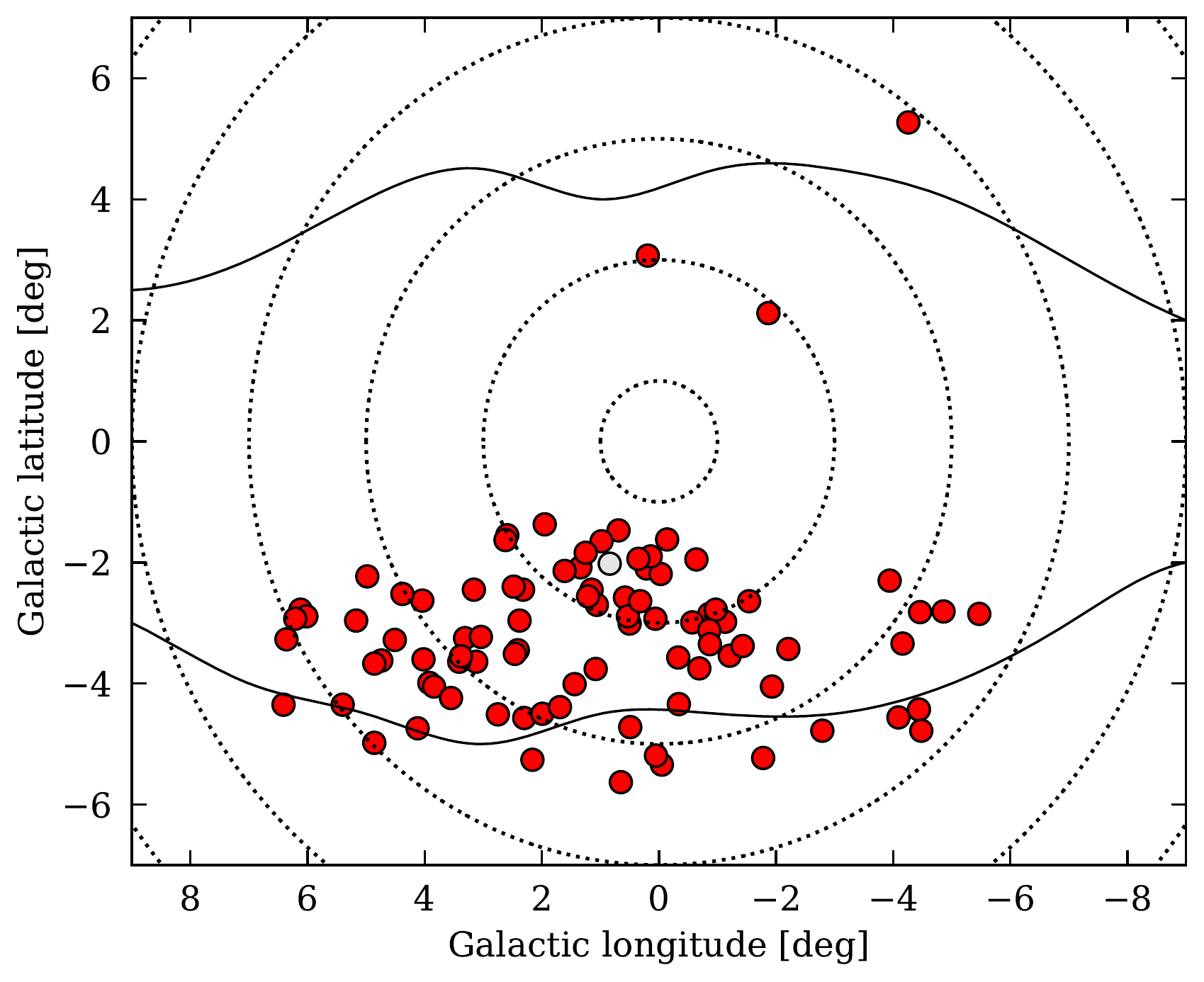}}
\caption{
Positions on the sky for the microlensed dwarf sample. The bulge contour lines based on observations with the COBE satellite are shown as solid lines \citep{weiland1994}. The dotted lines are concentric circles in steps of 2{\degr}. The star that might not be located within the bulge boundaries (OGLE-2013-BLG0911S) is marked by a grey circle (see \citealt{bensby2017} for further discussion).
\label{fig:glonglat}
}
\end{figure}
\begin{figure*}
\centering
\resizebox{0.95\hsize}{!}{
\includegraphics{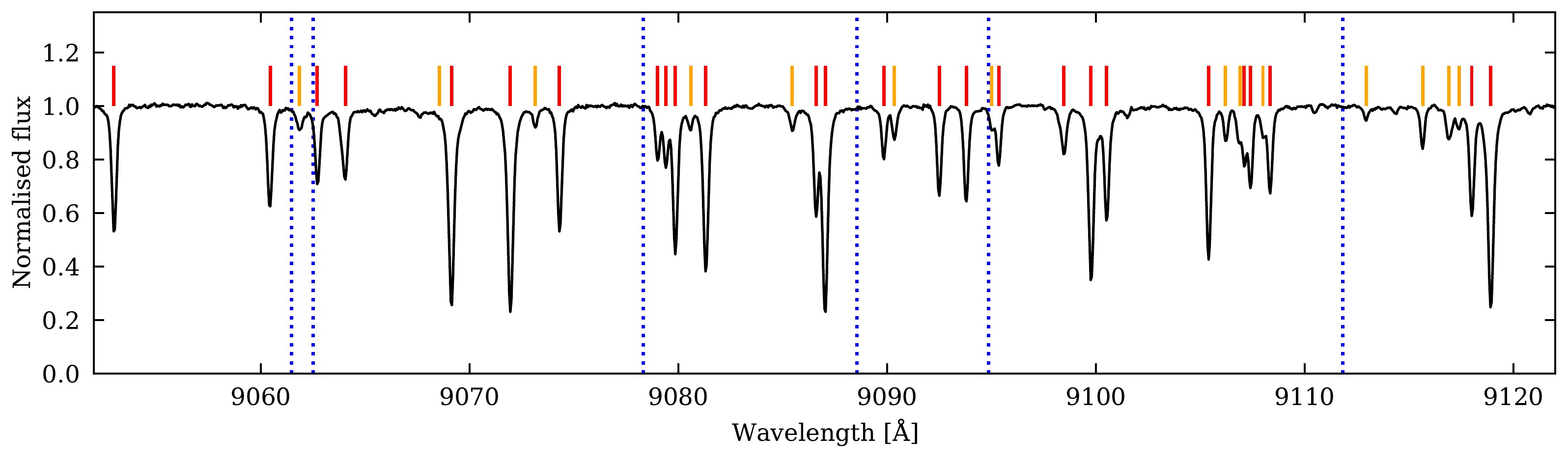}}
\caption{A spectrum of a rapidly rotating B star containing only telluric absorption lines.
The positions of the telluric lines are marked by red and orange lines (red lines deeper than orange ones, see Sect.~\ref{sec:telluric}), and the rest wavelengths of the six \ion{C}{i} lines are marked with dotted blue lines. Depending on the (geocentric) radial velocity of the target, the \ion{C}{i} lines will be either red- or blue-shifted and may, or may not, be contaminated by the telluric lines.
\label{fig:telluric}}
\end{figure*}

\begin{figure*}
\centering
\resizebox{0.95\hsize}{!}{
\includegraphics[viewport= 0 45 960 225,clip]{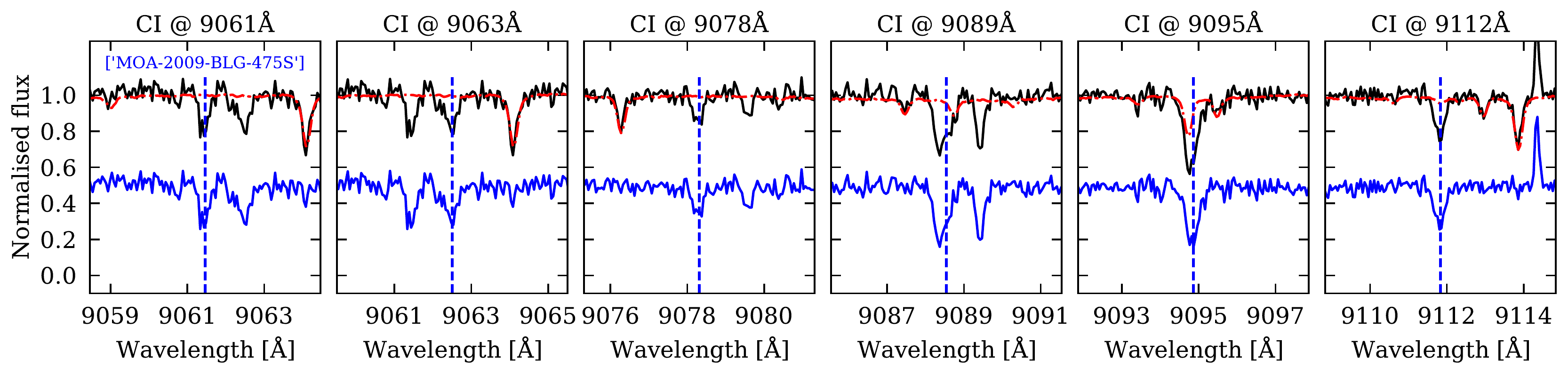}}
\resizebox{0.95\hsize}{!}{
\includegraphics[viewport= 0 0 960 205,clip]{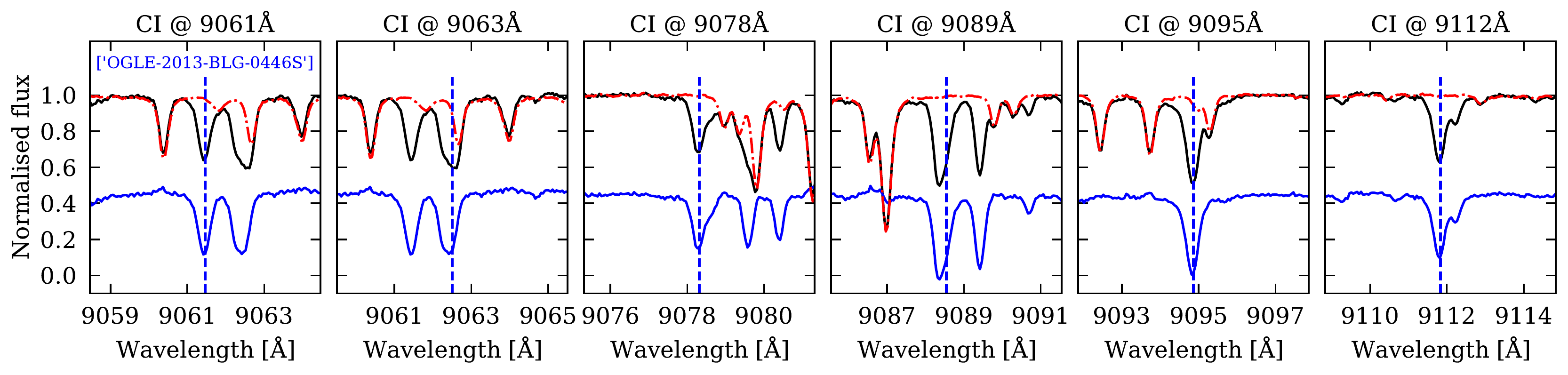}}
\caption{The \ion{C}{i} lines in the 9060-9120\,{\AA} region. The top panel shows an example for MOA-2009-BLG-493S and the bottom panel an example for OGLE-2013-BLG-0446S. The black lines represent the observed spectrum, the red dash-dotted line represent a telluric template based on several observations of rapidly rotating B stars, and the blue lines represent the observed spectrum (vertically shifted for graphical reasons), after removal of telluric features using the IRAF task {\sc telluric}. The vertical dashed lines mark the central wavelengths of the \ion{C}{i} lines. Note that the spectra have been shifted to rest wavelengths and that the telluric lines will contaminate different lines depending on the radial velocity of the star.
\label{fig:carbonlines}
}
\end{figure*}
\begin{figure}
\centering
\resizebox{\hsize}{!}{
\includegraphics{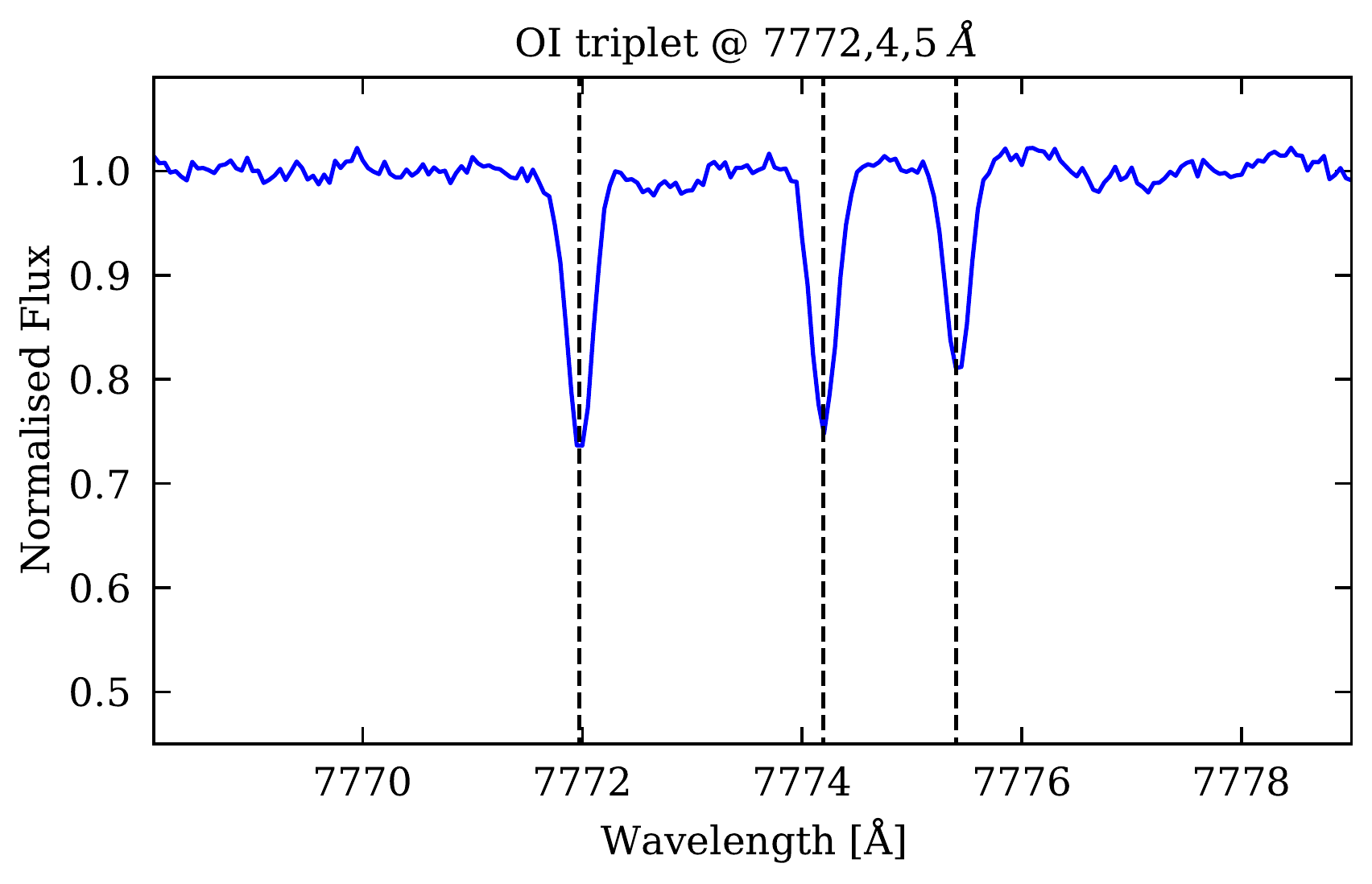}}
\caption{The \ion{O}{i} triplet lines at 7772--7775\,{\AA} that have been investigated in this study. The blue lines represents the observed spectrum of OGLE-2013-BLG-0446S. The vertical dashed lines mark the central wavelengths of the three oxygen lines.
\label{fig:oxygenlines}
}
\end{figure}

The stellar sample consists of 91 dwarf, turn-off, and subgiant stars in the Galactic bulge that were observed with high-resolution spectrographs during gravitational microlensing events. At the distance of the bulge it is extremely difficult to obtain good high-resolution spectra for such stars as they have apparent magnitudes around $V=18-21$. However, during microlensing events the brightnesses can increase by factors of several hundreds, making it possible to get high-resolution spectra suitable for detailed elemental abundance analysis with 2 hours, or shorter, exposures, using large telescopes such as the VLT, Magellan, or Keck. In the years 2009-2015 we were running a target-of-opportunity program with UVES at the VLT (PI: S. Feltzing), and accompanied with a dozen of observations with MIKE on Magellan, and HIRES on Keck, a total of 91 microlensing source stars towards the Galactic bulge were observed. For each target that was observed with UVES we also observed a rapidly rotating B star, resulting in a featureless spectrum only containing the telluric lines. This B star spectrum turns out to be crucial for the current analysis as the carbon lines around 9100\,Å that we utilise are located in a spectral region that is severely affected by telluric absorption lines.  The wavelength coverage of the spectra obtained with Keck does not include the carbon lines, and the spectra obtained with MIKE suffer {\bf severely} from fringing patterns in this spectral region, and also, the MIKE observations were not accompanied by observations of telluric standard stars. This means that carbon abundances will only be determined for those microlensed stars that were observed with UVES, that is, 70 stars out of the 91 stars in the sample. Oxygen abundances from the triplet lines at 7770\,Å will, however, be determined for all 91 stars. More details on the light curves for the microlensing events, the instruments that were used, and the quality of the obtained spectra can be found in \cite{bensby2017}.

Stellar parameters, stellar ages, and abundances for 12 elements (Na, Mg, Al, Si, Ca, Ti, Cr, Ni, Fe, Zn, Y, and Ba) were determined for all stars in \cite{bensby2017}. The methodology to determine stellar parameters and abundances from equivalent width measurements is identical to the study of 714 F and G dwarf stars in the Solar neighbourhood by \cite{bensby2014}. Briefly, standard 1-D plane-parallel model stellar atmospheres calculated with the Uppsala MARCS code  \citep{gustafsson1975,edvardsson1993,asplund1997} were used, and elemental abundances were calculated with the Uppsala EQWIDTH program using equivalent widths that were measured by hand using the IRAF\footnote{IRAF is distributed by the National Optical Astronomy Observatories, which are operated by the Association of Universities for Research in Astronomy, Inc., under cooperative agreement with the National Science Foundation \citep{tody1986,tody1993}.} task SPLOT. Effective temperatures were determined from excitation balance of abundances from \ion{Fe}{i} lines, surface gravities from ionisation balance between abundances from \ion{Fe}{i} and \ion{Fe}{ii} lines, and the microturbulence parameters by requiring that the abundances from \ion{Fe}{i} lines are independent of line strength. Line-by-line NLTE corrections from \cite{lind2012} were added to all iron abundances derived from \ion{Fe}{i} lines.

Stellar ages, masses, luminosities, absolute $I$ magnitudes ($M_I$), and colours ($V-I$) have been estimated from $Y^2$ isochrones \citep{demarque2004} by maximising probability distribution functions as described in \cite{bensby2011}. As shown in \cite{bensby2017} these age estimations are in very good agreement with Bayesian age methods such as the one by \cite{jorgensen2005}. \cite{valle2015} also validated the ages of the microlensed bulge dwarfs in \cite{bensby2013} using their method and found very good agreement.

Figure~\ref{fig:glonglat} shows the distribution of the sample in the plane of Galactic coordinates, and as can be seen most stars are contained within $-8\degr<l<8\degr$ and $-5\degr<b<-2\degr$. At a distance of 8\,kpc one degree in Galactic latitude is equivalent to 140\,pc which means that the great majority of the microlensed stars are located approximately $300-700$\,pc below the plane. This assumes that the stars are located at 8\,kpc, and if they are on the close side of the bulge they are closer to the plane while if they are on the far side they are also located farther from the plane. As we discussed in \cite{bensby2013,bensby2017} it is very unlikely that the source stars of these microlensing events are located outside the bulge region, and the geometry and microlensing likelihood favours them to be located on average slightly on the far side of the Galactic centre. Furthermore, the un-lensed magnitudes of the stars are what one can expect for turn-off and subgiant stars at the distance of the Galactic bulge. This, together with the distance-independent spectroscopic determination of the stellar parameters that shows that they are turn-off and subgiant stars, further strengthen that the stars are located in the bulge region, that is, within two to three kiloparsecs of the Galactic centre. One exception is OGLE-2013-BLG-0911S that, based on microlensing arguments, likely is located in the disk on the close edge of the Bulge (see discussion in \citealt{bensby2017}).

\begin{figure*}
\centering
\resizebox{\hsize}{!}{
\includegraphics[viewport= 0 0 750 500,clip]{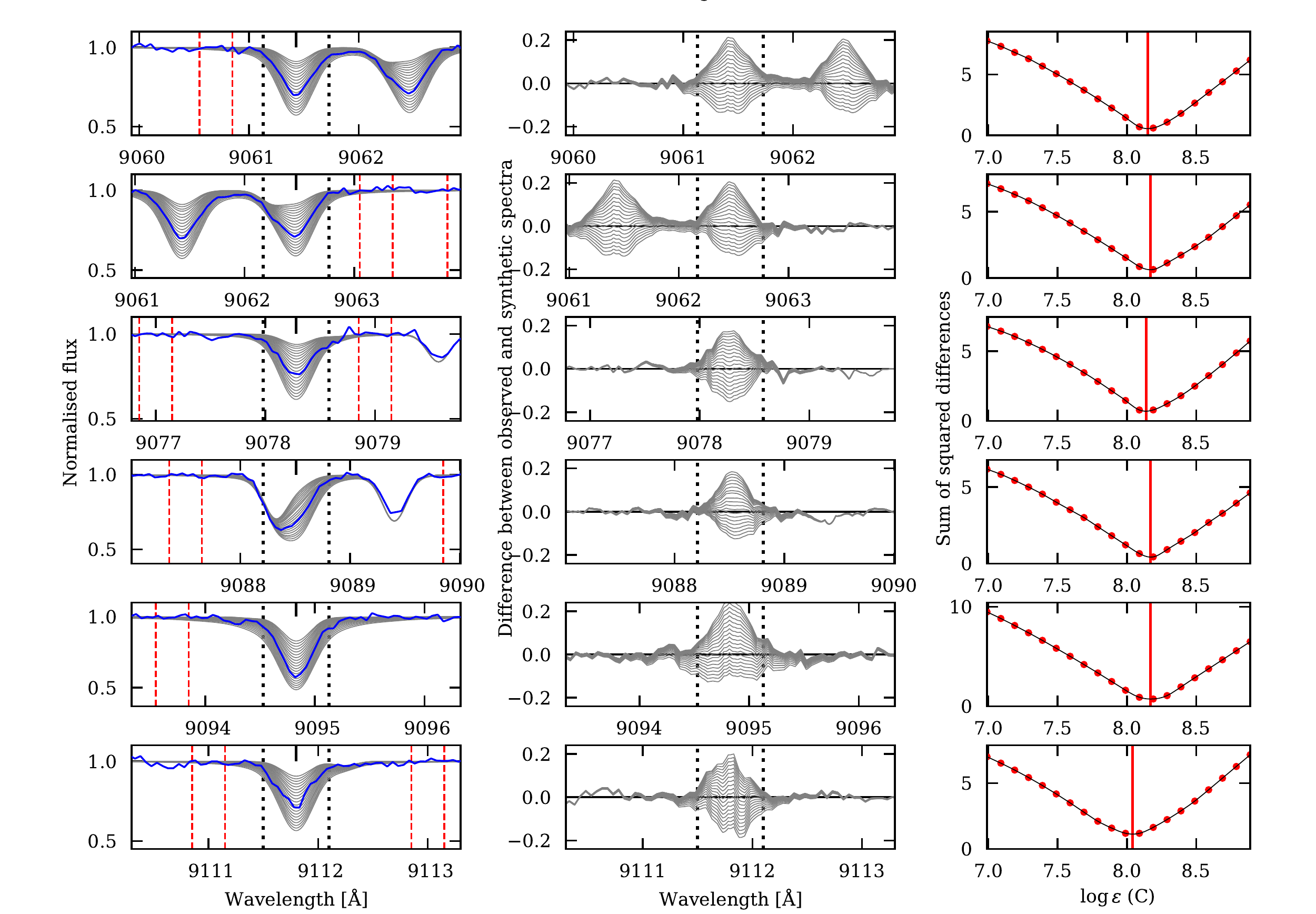}}
\resizebox{\hsize}{!}{
\includegraphics[viewport= 0 235 750 500,clip]{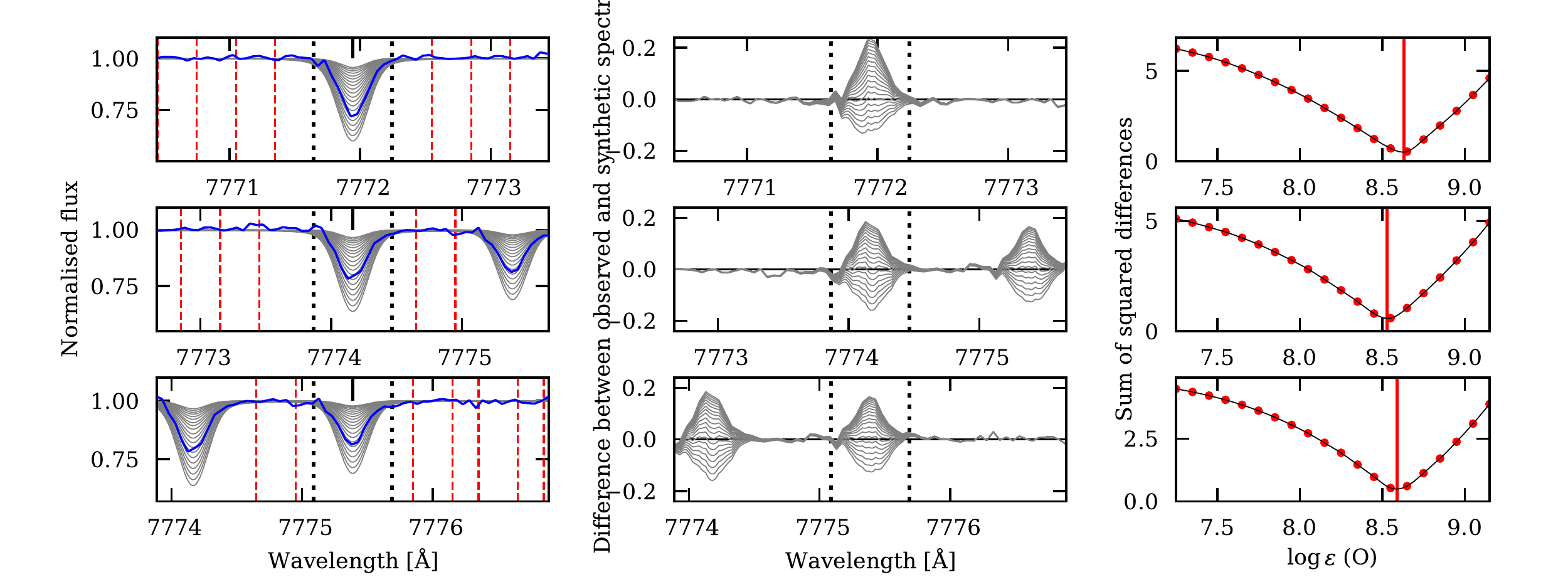}}
\caption{Example showing the determination of carbon abundances from the six different \ion{C}{I} lines (plots in the top six rows) and the three \ion{O}{i} lines (plots in the three bottom rows). The plots in the left column shows the observed spectrum as red lines and 20 synthetic spectra with carbon abundances in steps of 0.1\,dex as black lines. The plots in the middle column show the differences between the observed and synthetic spectra. The plots in the right column shows the sum of the squared differences between observed and synthetic spectra within $\pm0.3$ of the central wavelengths of the \ion{C}{I} and \ion{O}{i} lines (marked by black vertical dotted lines in the plots in the left column). The vertical red solid line shows the minimum of the $\chi^2$-values for which the best fitting carbon and oxygen abundances are selected. For each line we have done a local normalisation of the observed spectrum, and the selected continuum regions are marked by dotted red vertical in the plots in the left column.
\label{fig:carbonfits}
}
\end{figure*}

\section{Abundance analysis}
\label{sec:analysis}

\subsection{Spectral lines and atomic data}

To determine carbon abundances we use six \ion{C}{i} lines that are located in the near-infrared part of the spectrum in the range 9060 to 9120\,{\AA}. These \ion{C}{i} lines are located in a region that is contaminated by telluric lines, meaning that before an abundance analysis can be performed, the telluric features must be removed from the bulge dwarf spectra. Therefore, for each target that was observed we also, during the same night, observed a rapidly rotating B star, resulting in a featureless spectrum only containing the telluric lines (see Fig.~\ref{fig:telluric}). The red lines mark telluric lines with a line depth greater than about $15$\% of the continuum, and the orange lines mark those with a smaller depth. Also indicated by blue lines are the locations of the six \ion{C}{i} lines. Depending on the radial velocity of the star, the carbon lines may, or may not, be blended by the telluric lines.

Using the IRAF task TELLURIC this spectrum can then be scaled and shifted slightly before it is used to divide out the telluric features in the observed spectrum of the bulge star. Figure~\ref{fig:carbonlines} shows examples of the results of the telluric removal procedure for two of our targets. As can be seen the procedure is efficient, generally leaving negligible traces of the telluric lines, allowing us to make use of even those \ion{C}{i} lines that were directly aligned with the telluric absorption features. 

The two \ion{C}{i} lines at 9063\,{\AA} and 9089\,{\AA} have asymmetric line profiles, and they are also shifted slightly off-centre relative to the reference wavelengths.  Figure~\ref{fig:carbonlines} shows that this cannot be due to a poor removal of the telluric lines. The asymmetries and offsets of these two lines are caused by the presence of other blending spectral lines. In the case of the 9062\,{\AA} line there are several blending lines but the main contributor is the \ion{Fe}{i} line at  9062.2392\,{\AA}, and for the 9088\,{\AA} \ion{C}{i} line there is an \ion{Fe}{i} line at 9088.3177\,{\AA}. This makes the analysis of these \ion{C}{i} lines challenging as the absorption profile usually is dominated by the blending line. However, we will keep them for the abundance analysis and then evaluate their usability (see Sect.~\ref{sec:telluric}).

We also considered the \ion{C}{i} lines at 7111 and 7113\,{\AA}. But because these lines are weak they could rarely be discerned from the continuum noise, given the sometimes relatively low $S/N$ ratios, we decided not to use them.

The infrared \ion{O}{i} triplet lines at 7772--7775\,{\AA} were used to determine oxygen abundances. In solar-type stars these lines are usually strong and clean as they are located in a wavelength region free from other blending lines and telluric lines (see example in Fig.~\ref{fig:oxygenlines}). Another option would be to use the forbidden \ion{O}{i} line at 6300\,{\AA}, but unfortunately that line is too weak to be used in the microlensed bulge dwarf spectra.

Table~\ref{tab:atomdata} lists the atomic data for the C and O lines that were analysed. The atomic line data for the \ion{C}{i} and \ion{O}{i} lines, as well as the other spectral features in the region, were gathered from the Vienna Atomic Line Database   \citep[VALD,][]{vald_1,vald_2,vald_3,vald_4,vald_5,vald_6}. 

\begin{table}[t]
\centering
\setlength{\tabcolsep}{1.5mm}
\caption{
The carbon and oxygen  lines investigated in this study, and the solar abundances that we determine from each line.\tablefootmark{$\dagger$}
\label{tab:atomdata}
}
\tiny
\begin{tabular}{ccrccc}
\hline\hline
\noalign{\smallskip}
Elem.   & $\lambda$         & $\log (gf)$ & $\chi_{\rm l}$  &  $\log \epsilon (X)$  & Reference \\
        & ({\AA})             &             & (eV)          &   NLTE                & \\
\noalign{\smallskip}
\hline
\noalign{\smallskip}
\ion{O}{i} & 7771.944 & $ 0.369$ & 9.1461 & 8.61   & H91, BZ92 \\
\ion{O}{i} & 7774.166 & $ 0.223$ & 9.1461 & 8.64   & H91, BZ92 \\
\ion{O}{i} & 7775.388 & $ 0.002$ & 9.1461 & 8.63   & H91, BZ92 \\
\ion{C}{i} & 9061.430 & $-0.347$ & 7.4828 & 8.37   & HK17, W96, H93, NS84  \\
\ion{C}{i} & 9062.470 & $-0.455$ & 7.4804 & 8.40   & HK17, W96, H93, NS84  \\
\ion{C}{i} & 9078.280 & $-0.581$ & 7.4828 & 8.45   & HK17, W96, H93, NS84  \\
\ion{C}{i} & 9088.510 & $.0.430$ & 7.4828 & 8.41   & HK17, W96, H93, NS84  \\
\ion{C}{i} & 9094.830 & $ 0.151$ & 7.4878 & 8.48   & HK17, W96, H93, NS84  \\
\ion{C}{i} & 9111.800 & $-0.297$ & 7.4878 & 8.43   & HK17, W96, H93, NS84  \\
\noalign{\smallskip}
\hline
\end{tabular}
\tablefoot{
\tablefoottext{$\dagger$}{
For each line we give the wavelength (air), the $\log gf$ value, the lower excitation energy ($\chi_{\rm l}$), the absolute abundance ($\log \epsilon (X)$), and the references for the oscillator strengths: H91 = \cite{hibbert1991}, BZ92=\cite{biemont1992}, H17=\cite{haris2017}, W96=\cite{wiese1996}, H93=\cite{hibbert1993}, NS84=\cite{nussbaumer1994}. 
}}
\end{table}

\subsection{Line synthesis and NLTE corrections}

The abundance analysis was done through a $\chi^2$-minimisation between the observed spectrum and a synthetic spectrum.
The synthetic spectra are calculated with pySME which is the python implementation of the spectroscopy made easy (SME) software \citep{valenti1996,piskunov2017}. 

To calculate a synthetic spectrum pySME needs atomic data (see above), stellar parameters, broadening parameters, and model stellar atmospheres. For the latter we use the MARCS model atmospheres \citep{gustafsson2008}, while $\teff$, $\log g$, [Fe/H], and microturbulence are taken from our detailed analysis of the stars in \cite{bensby2017}. 

In addition to atomic line broadening, the observed line profile is broadened by the instrument, the line-of-sight component of the stellar rotation ($v_{\rm rot} \cdot \sin i$), and small- and large-scale motions in the stellar atmosphere (microturbulence, $\xi_{\rm t}$, and macroturbulence, $v_{\rm macro}$, respectively). The instrument broadening is set by the resolving power of the spectrograph ($R$) and is treated with a Gaussian profile, while $v_{\rm rot} \cdot \sin i$ and $v_{\rm macro}$ are jointly accounted for with a radial-tangential (RAD-TAN) profile. To determine the RAD-TAN broadening other, unblended, relatively weak, and unsaturated, \ion{Fe}{i} lines located at 6065, 6546, and 6678\,{\AA} were used. The C and O abundances for the different \ion{C}{i} and \ion{O}{i} lines were then determined through a simple $\chi^{2}$-minimisation routine that is illustrated in Fig.~\ref{fig:carbonfits}.

All of the \ion{C}{i} and \ion{O}{I} lines that are used in this study are sensitive to departures from the assumption of local thermodynamic equilibrium (LTE), and large non-LTE (NLTE) corrections are needed to achieve reliable abundances \citep[e.g.][]{asplund2005araa,amarsi2019b}. For the oxygen triplet we have in the previous papers \citep{bensby2009,bensby2011,bensby2013} applied the empirical correction formula from \cite{bensby2004}, which studied the forbidden oxygen line at 6300\,{\AA} that is insensitive to departures from LTE \citep[e.g.][]{asplund2004}. However, that formula was based on a smaller sample of F and G dwarf stars that do not cover the full range of stellar parameters that are spanned by the microlensed dwarf sample. Instead we now make use of the tables of NLTE departure coefficients, for both C and O, from \cite{amarsi2020b} that are implemented directly into pySME.

\begin{table}[t]
\centering
\caption{Description of the online table that gives
the carbon and oxygen abundances (NLTE) and their associated uncertainties.
\label{tab:abundances}
}
\tiny
\begin{tabular}{llll}
\hline\hline
\noalign{\smallskip}
Column   & Parameter & Unit & Description\\
\noalign{\smallskip}
\hline
\noalign{\smallskip}
(1) & Star    & & MACHO, OGLE, or MOA id.\\
(2) & $\rm [Fe/H]$  & Sun & Abundance ratio normalised to Sun \\
(3) & $\rm [C/H]$   & Sun & " \\
(4) & $\rm [O/H]$   & Sun & " \\
(5) & $\rm\sigma[Fe/H]$   & dex & Uncertainty in abundance ratio\\
(6) & $\rm\sigma[C/H]$   & dex & " \\
(7) & $\rm\sigma[O/H]$   & dex & " \\
(8) & $\rm\sigma[O/Fe]$   & dex & " \\
(9) & $\rm\sigma[C/Fe]$   & dex & " \\
(10) & $\rm\sigma[C/O]$    & dex & " \\
(11) & A(O7772)   & dex & Absolute abundance from line \\
(12) & A(O7774)   & dex & " \\
(13) & A(O7775)   & dex & " \\
(14) & A(C9061)   & dex & " \\
(15) & A(C9063)   & dex & " \\
(16) & A(C9078)   & dex & " \\
(17) & A(C9089)   & dex & " \\
(18) & A(C9094)   & dex & " \\
(19) & A(C9111)   & dex & " \\
(20) & Flag(9061)  &     & Telluric line flag\tablefootmark{$\dagger$} \\
(21) & Flag(9063)  &     & " \\
(22) & Flag(9078)  &     & " \\
(23) & Flag(9089)  &     & " \\
(24) & Flag(9094)  &     & " \\
(25) & Flag((9111)  &     & " \\

\noalign{\smallskip}
\hline
\end{tabular}
\tablefoot{
\tablefoottext{$\dagger$}{
For each line we give a flag (0, 1, or 2) depending on whether the spectral line is close to a telluric absorption line or not: $0 =$ not affected; $1=$ closer than 0.3\,{\AA} to a weak telluric line (depth smaller than 15\,\% o fthe continuum level); $2=$ closer than 0.3\,{\AA} to a strong telluric line (depth greater than 15\,\% of the continuum level).
}}
\end{table}

Abundances of iron, carbon and oxygen for all stars and all lines are given in Table~\ref{tab:abundances}. Also included in Table~\ref{tab:abundances} are flags that indicate whether the individual \ion{C}{i} lines are close to a weak or a strong telluric line or not at all.

\subsection{Solar analysis}

The derived abundances will be normalised to the Sun on a line-by-line basis. The solar analysis was carried out on spectra obtained through observations of the asteroids Vesta and Ceres. This means that rather than taking a standard value for the Sun we determine our own oxygen and carbon abundances for each and every line that we analyse and use those to normalise the individual line abundances. In that way we will minimise systematic uncertainties that are likely to arise because of the analysis methods and atomic data that are used. Our analysis is therefore truly differential to the Sun. Table~\ref{tab:atomdata} gives our carbon and oxygen results for the Sun, and as can be seen they are internally consistent and compare well with the abundances derived by, for instance, \cite{amarsi2019a}, who performed a 3D NLTE line formation analysis of neutral carbon in the Sun ($\rm A(C)=8.44$), and \cite{amarsi2018}, who found $\rm A(O)=8.69$ from the \ion{O}{i} triplet lines, again using 3D NLTE models.

The NLTE corrected solar abundances listed in Table~\ref{tab:atomdata} will be used to normalise the abundances on a line-by-line basis.

\subsection{Abundances from individual lines}
\label{sec:telluric}

\begin{figure}
\centering
\resizebox{\hsize}{!}{
\includegraphics{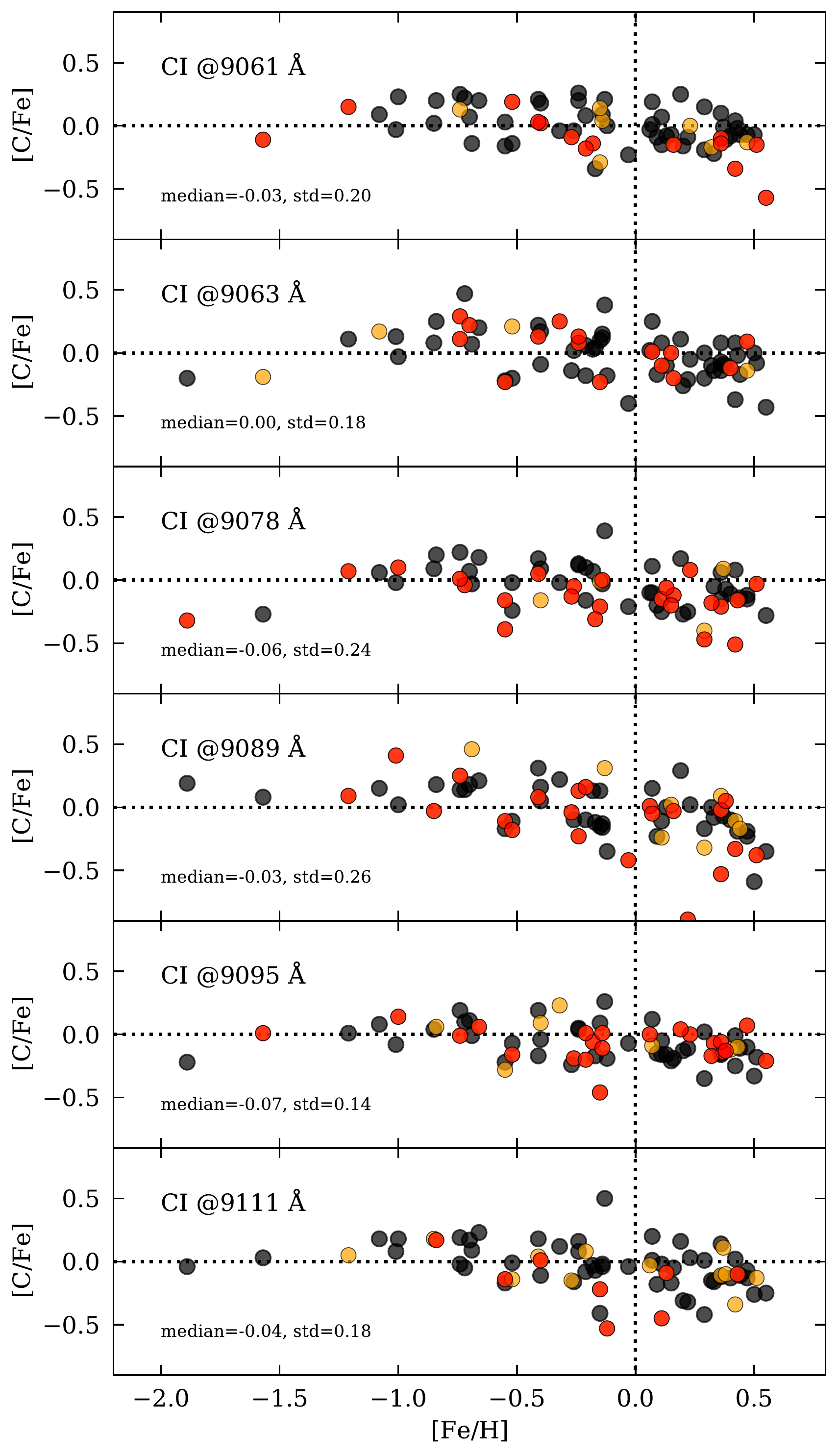}}
\caption{
[C/Fe]-[Fe/H] trend based on individual \ion{C}{i} lines (as indicated in the plots). Black circles indicate that the \ion{C}{i} line in question have not been affected by a telluric line, orange circles mean that the line in question has been located closer than 0.3\,{\AA} to a telluric line with a depth weaker than 15\,\% of the continuum, and the red circles mean that the line in question has been located closer  than 0.3\,{\AA} to one of the stronger telluric lines. The telluric lines are marked in Fig.~\ref{fig:telluric}. In each plot we have also indicated the median [C/Fe] value and the dispersion in [C/Fe].
\label{fig:cfe_lines}}
\end{figure}
\begin{figure}
\centering
\resizebox{\hsize}{!}{
\includegraphics{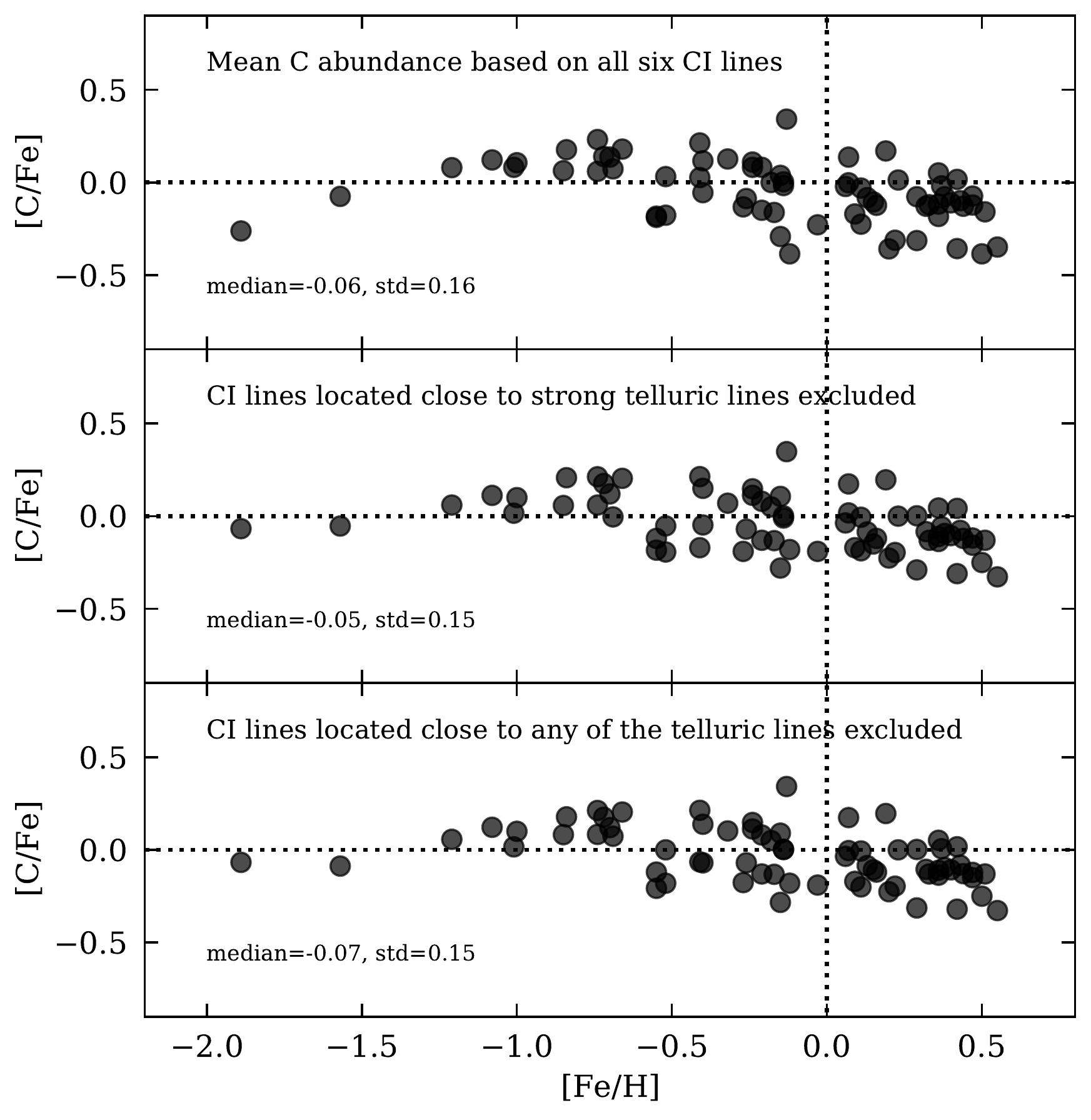}}
\caption{
The [C/Fe]-[Fe/H] trends where the carbon abundances are based on the mean abundance coming from the different \ion{C}{i} lines. In the top plot all \ion{C}{i} lines are included in the calculation of the mean C abundance for each star; in the middle plot \ion{C}{i} lines located close to strong telluric lines (red marked lines in Fig.~\ref{fig:telluric}) were excluded when calculating the mean C abundance for each star; and in the bottom plot all \ion{C}{i} lines located close to a telluric line (orange and red lines in Fig.~\ref{fig:telluric} and orange and red points in Fig.~\ref{fig:cfe_lines}) were excluded in the calculation of the mean C abundance for each star. In each plot we have also indicated the median [C/Fe] value and the dispersion in [C/Fe].
\label{fig:cfe_lines_mean}}
\end{figure}

Figure~\ref{fig:cfe_lines} shows the [C/Fe]-[Fe/H] trends based on carbon abundances from the individual \ion{C}{i} lines. Here we have marked the stars where the line centre of the \ion{C}{i} line that the carbon abundance is based on falls within $\pm0.3$\,{\AA} of a telluric line (orange marker if it is one of the weaker telluric lines and a red marker if it is one of the stronger telluric lines, as shown in Fig.~\ref{fig:telluric}). In each plot we have also indicated the median [C/Fe] value and the dispersion in [C/Fe]. First we note that all six \ion{C}{i} lines are able to give carbon abundances that result in very similar $\rm [C/Fe]-[Fe/H]$ abundance trends, even the two lines at 9063\,{\AA} and 9089\,{\AA} that were blended by other \ion{Fe}{i} lines in the stellar spectrum. We also note that there are no particular trends for those stars whose lines originally were originally located close to the telluric lines. This is further illustrated in Fig.~\ref{fig:cfe_lines_mean}, which shows the $\rm [C/Fe]-[Fe/H]$ trends based on the average abundance from all six \ion{C}{i} lines, with the top plot including all lines, the middle plot those lines that were unaffected by telluric lines or close to one of the weaker telluric lines (orange lines in Fig.~\ref{fig:telluric}), while the bottom plot only contain the lines that were completely unaffected by telluric lines. The median [C/Fe] values and the dispersions in [C/Fe] are very similar in all three cases (the numbers are given in Fig.~\ref{fig:telluric}). The three plots are very similar, confirming that the removal of the telluric lines from the microlensed dwarf star spectra has been successful.

As there is no clear indication that any of the six \ion{C}{i} lines is producing unusual C abundances the discussion and figures will from now on be based on the average C abundances inferred from all six lines (whenever available).

\subsection{Error analysis}

Random uncertainties in the derived carbon and oxygen abundances are estimated by taking the uncertainties in the stellar parameters ($\teff$, $\log g$, [Fe/H], and $\xi_{\rm t}$) into consideration. The changes in the abundances that are acquired by changing the stellar parameters are then added in quadrature (this is a relatively easy, and commonly used, way to estimate uncertainties, but assumes that the uncertainties in the stellar parameters are un-correlated, which might not be completely true). The uncertainties are given in Table~\ref{tab:abundances}, and also shown in Fig.~\ref{fig:ofecfe}.

Systematic uncertainties should be minimised as we are doing a strictly differential analysis to the Sun. The stars have quite similar parameters relative to each other and relative to the Sun so differential systematic errors due to, for example, 3D, non-LTE, stellar parameters, line broadening should be small (and atomic data errors largely cancel). By confirming that the different spectral lines give consistent abundances and very similar abundance trends (see Fig.~\ref{fig:cfe_lines}), we also are confident that there should be no major systematic offsets.

\section{Results}
\label{sec:results}

\begin{figure*}
\centering
\resizebox{0.65\hsize}{!}{
\includegraphics{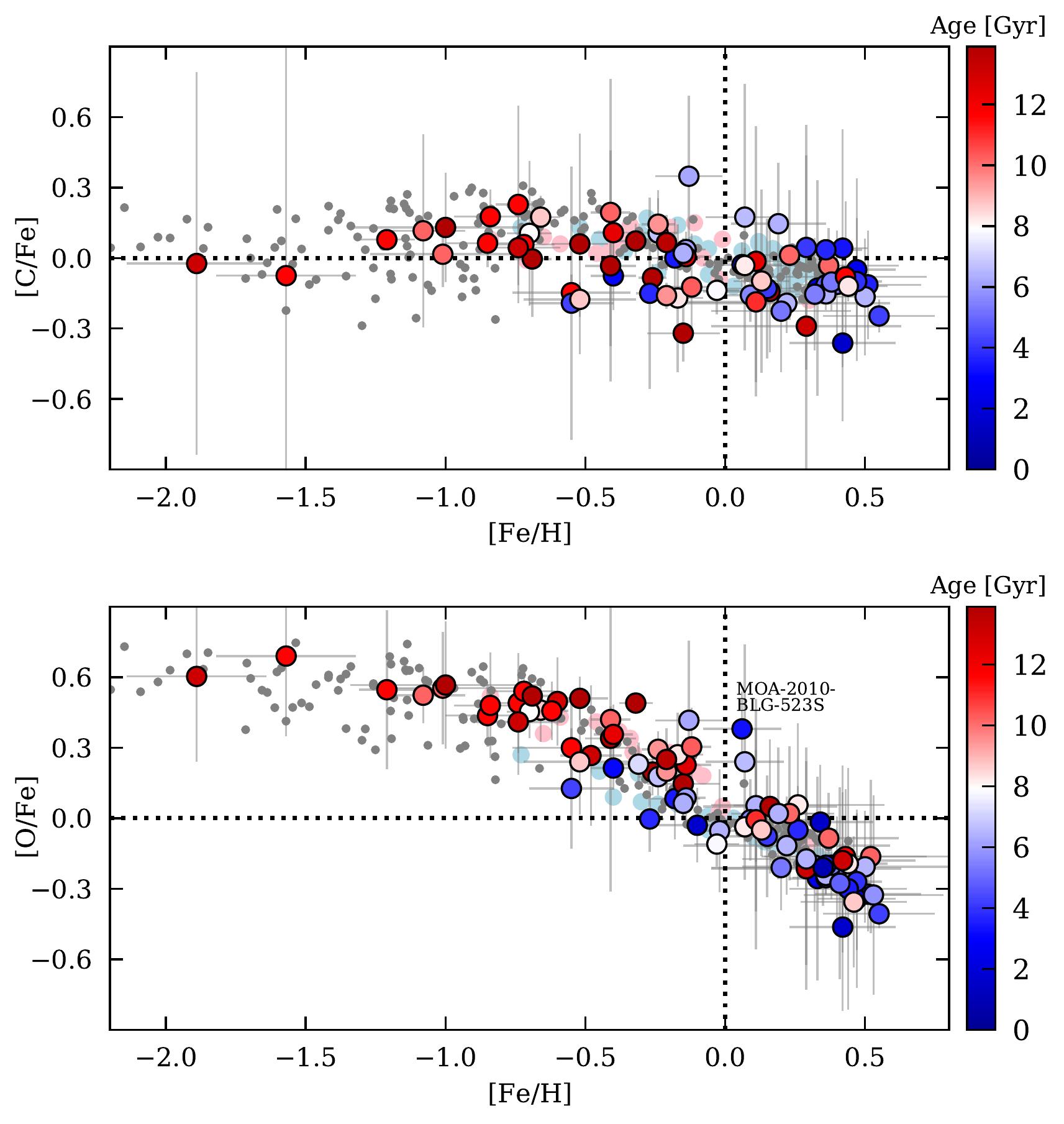}}
\resizebox{0.65\hsize}{!}{
\includegraphics{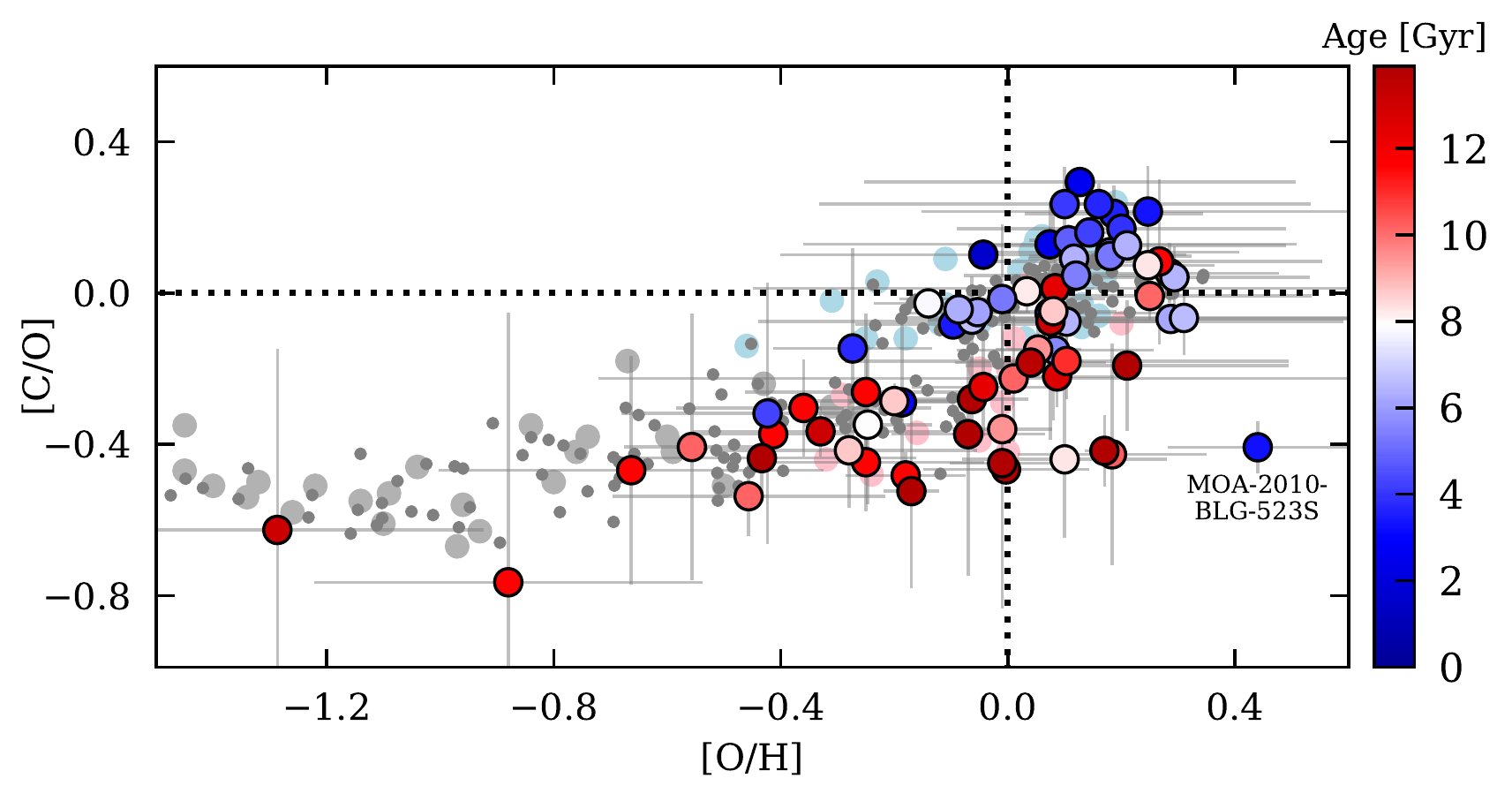}}
\caption{[C/Fe] versus [Fe/H] (top panel), [O/Fe] versus [Fe/H] (middle panel), and [C/O] versus [O/H] (bottom panel)  for the microlensed bulge dwarf sample. The bulge stars have been colour-coded based on their ages, according to the colour bar on the right-hand side. Light-red circles and light-blue dots are solar neighbourhood thick and thin disk dwarf stars, respectively (oxygen based on the forbidden [\ion{O}{i}] line at 6300\,{\AA} from \citealt{bensby2004,bensby2005} and carbon based on the forbidden [\ion{C}{i}] line at 8727\,{\AA} from \citealt{bensby2006}). Large grey circles in the [C/H]-[O/H] plot are halo stars from \cite{akerman2004}, and small grey dots (in all plots) are the disk and halo stars (1D/NLTE results) from \cite{amarsi2019b}.
In the oxygen plots the positions of MOA-2010-BLG-523S has been marked. In the carbon plot at the top, MOA-2010-BLG-523S has a [C/Fe] value very close to solar. 
\label{fig:ofecfe}}
\end{figure*}

Figure~\ref{fig:ofecfe} shows the $\rm [C/Fe]-[Fe/H]$, $\rm [O/Fe]-[Fe/H]$, and $\rm [C/O]-[O/H]$ abundance trends for the microlensed bulge dwarf sample. For comparison purposes the plots also include stars representing the Galactic thin and thick disks where the abundances are based on the forbidden \ion{C}{i} and \ion{O}{i} lines at 8727\,{\AA} and 6300\,{\AA}, respectively \citep{bensby2004,bensby2005,bensby2006}. These forbidden lines are not sensitive to departures from the assumption of LTE \citep[e.g.][]{asplund2004,asplund2005}. Also included is the sample from \citealt{amarsi2019b} that presented a 1D/3D LTE/NLTE analysis of C and O for 187 nearby disk and halo stars (the 1D/NLTE results from \citealt{amarsi2019b} are shown here). In the $\rm [C/O]-[O/H]$ plot we further include the halo sample from \cite{akerman2004}, based on NLTE corrected abundances from the same C and O lines analysed in this study. These disk and halo trends compare well with the \cite{amarsi2019b} results.

For the microlensed dwarf stars, as well as the comparison samples, the [C/Fe] ratios tend to be slightly enhanced compared to the Sun for metallicities up to $\rm [Fe/H]\approx0$ and thereafter they decline towards higher [Fe/H]. A slight difference in the details is that \cite{amarsi2019b} see a separation in the [C/Fe]-[Fe/H] diagram  between thin and thick disk stars, while \cite{bensby2006} do not see that. An explanation could be that the two studies use different ways of defining thin and thick disk stars: \cite{amarsi2019b} uses chemical criteria (as defined by \citealt{adibekyan2013}), while \cite{bensby2006} uses kinematical criteria. In this study we use age criteria, as advocated for in \cite{bensby2014}, where thin disk stars are likely to be younger than 8\,Gyr, and thick disk stars are likely to be older than 8\,Gyr. It should be noted that the overall appearance of the $\rm [C/Fe]-[Fe/H]$ trend in the disk is very similar between the two studies. This also holds for the $\rm [O/Fe]-[Fe/H]$ trend where \cite{bensby2004,bensby2005} agree well with \cite{amarsi2019b}, that is, a clear distinction between the thin and thick disks at sub-solar [Fe/H] and a continued, rather steep, decline in [O/Fe] towards higher [Fe/H].  

\subsection{Carbon in the bulge}

The evolution of [C/Fe] varies essentially in lockstep with [Fe/H] for the microlensed bulge dwarf stars. There is a slight tendency of a carbon over-abundance at sub-solar metallicities and a carbon under-abundance at super-solar metallicities, that also can be interpreted as a shallow decline in [C/Fe] with [Fe/H]. The [C/Fe] trend for the stars younger than about 8\,Gyr follow smoothly upon the trend for the stars older than about 8\,Gyr, and there is no indication of 
a separation in [C/Fe] between the stars in the two age groups. Comparisons to the local thin and thick disk samples from \cite{bensby2006} to the $<8\,$Gyr and $>8\,$ Gyr subsamples show that the microlensed bulge dwarf [C/Fe] trends are very similar to what is observed in the nearby Galactic thin and thick disks.

\subsection{Oxygen in the bulge}

The $\rm [O/Fe]-[Fe/H]$ trend for the microlensed dwarf stars shows the typical signature abundance trend seen for the other $\alpha$-elements \citep{bensby2017}, that is, an enhanced [O/Fe] ratio that declines towards solar metallicities for the stars older than about 8\,Gyr, and a more shallow separated decline for the stars younger than about 8\,Gyr.  At super-solar metallicities, the $\rm [O/Fe]-[Fe/H]$ trend continues to decrease, in contrast to the other $\alpha$-elements, but in agreement with the [O/Fe] trend in the local disk \citep[as also seen in, e.g.,][but for a much smaller sample]{alvesbrito2010}.

The agreement between the bulge and local disk trends are striking, but with a reservation that the bulge [O/Fe] abundance ratios might be slightly more enhanced than what is seen in the local thick disk. Although it could not be statistically verified, in \cite{bensby2017} we suggested that this apparent enhancement could be because the position of the ``knee'' in the bulge is located at a metallicity that is about 0.1\,dex higher than in the thick disk. On the other hand, \cite{griffith2021} conclude that the differences between the bulge and local disk stars seen in several studies are due to sampling effects with bulge samples generally containing more evolved stars, and if this bias is compensated for, the differences vanish. Also \cite{zasowski2019} find that the position of the knee in the $\alpha$-element abundance trends is constant with galactocentric radius in the Galactic disk.

\subsection{Carbon-over-oxygen in the bulge}
\label{sec:results_co}

The bottom plot in Fig.~\ref{fig:ofecfe} shows the $\rm [C/O]-[O/H]$ abundance plane for the microlensed bulge dwarf stars, and it shows how the old and young stars in the bulge divide into the two distinct and well-separated sequences as outlined by the solar neighbourhood thin and thick disks. The metal-poor end of the sample aligns with the [C/O] ratios observed in the stellar halo samples by \cite{akerman2004} and \cite{amarsi2019b}.

\begin{figure}
\centering
\resizebox{\hsize}{!}{
\includegraphics{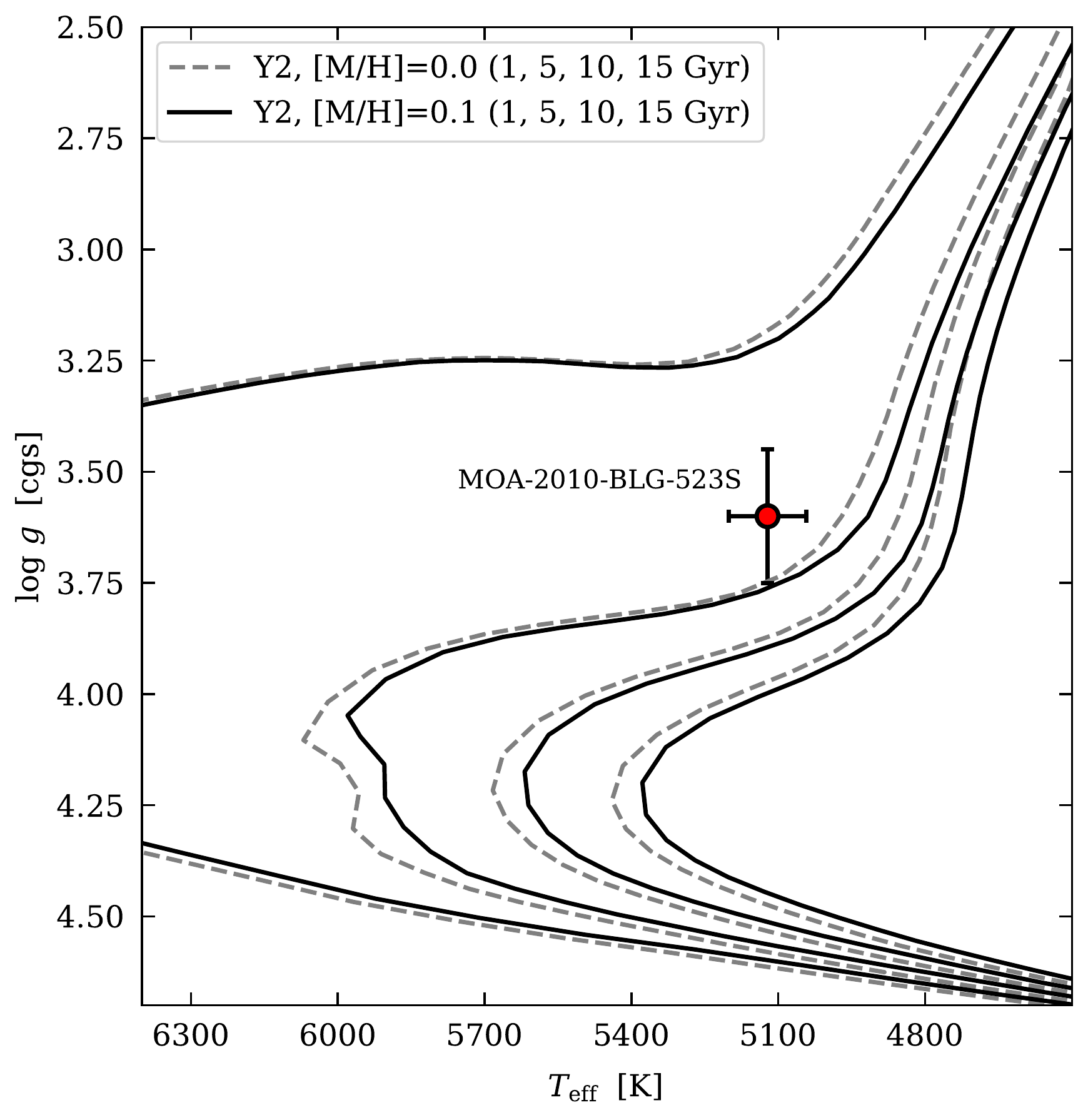}}
\caption{The position of MOA-2010-BLG-523S in the HR diagram. Two sets of isochrones from \cite{demarque2004} are plotted with metallicities as indicated in the figure. For each set of isochrones the ages of 1, 5, 10, and 15\,Gyr are shown. If the star is indeed an RS CVn star its metallicity is probably too low by up to 0.1\,dex, and its effective temperature too low by a few hundred degrees, while its surface gravity is likely to be unchanged. As can be seen that change has essentially had no effect on its estimated age.
\label{fig:hr}}
\end{figure}

\subsection{Outliers}
\label{sec:outliers}

Even though the overall carbon and oxygen abundance trends shown in Fig.~\ref{fig:ofecfe} are well-defined and tend to follow what is observed locally in the Galactic disk, there are a few stars that deviate. In particular there is one star in the $\rm [C/O]-[O/H]$ plot (MOA-2010-BLG-523S with $\rm [C/O]=-0.39$ at $\rm [O/H]=0.44$), and perhaps three stars (marked as younger than 8\,Gyr) with high [O/Fe] values around solar metallicities in the $\rm [O/Fe]-[Fe/H]$ plot. We have checked the analysis of those stars extra carefully, in particular, whether their spectra have unusually low SNR, whether the fitting of the synthetic spectra are bad, whether they have stellar parameter uncertainties that are unusually high, but no such indications were seen.

Regarding  MOA-2010-BLG-523S (with $\rm [C/O]=-0.39$ at $\rm [O/H]=0.44$, and $\rm [O/Fe]=+0.38$ and $\rm [C/Fe]=-0.01$ at $\rm [Fe/H]=0.06$) we note that this star actually was deemed by \cite{gould2013} to be an RS CVn star in the bulge. If this is so, it might be that the measured abundance ratios are affected, and in particular the NLTE effects  of the abundances from the \ion{O}{i} triplet lines are very large, that could lead to severely overestimated photospheric oxygen abundances  of up to 0.5-1.5\,dex \citep{morel2006}. Thus this could explain the very high [O/H] ratio leading to the very low [C/O] ratio. In the $\rm [O/Fe]-[Fe/H]$ and $\rm [C/O]-[O/H]$ plots in Fig.~\ref{fig:ofecfe} MOA-2010-BLG-523S has been marked out but not in the $\rm [C/Fe]-[Fe/H]$ plot as its [C/Fe] value does not stand out compared to the other stars. 

If MOA-2010-BLG-523S indeed is a RS CVn star, and depending on how high its activity index is, its stellar parameters might also have been affected. According to \cite{spina2020} it is mainly the microturbulence parameter that is affected, and overestimated, which would lead to too low [Fe/H] values of up to 0.1\,dex. The surface gravity appears essentially unaffected while the effective temperature could be underestimated by up to a few hundred degrees for stars that have high activity indices. We do not know the actual activity index of MOA-2010-BLG-523S, but as was pointed out in \cite{gould2013} its microturbulence deviates from the other stars in the sample, which in turn could mean that its temperature also could be underestimated. Figure~\ref{fig:hr} shows MOA-2010-BLG-523S in an HR diagram with matching isochrones over-plotted.  A higher metallicity of up to 0.1\,dex and a higher temperature of up to a few hundred degrees, would probably not change the estimated age of MOA-2010-BLG-523S by much as it will only move horizontally to the left in the HR diagram. Hence it is clear that this is still a young star located in the Galactic bulge.

\section{Discussion}
\label{sec:discussion}

\begin{figure*}
\centering
\resizebox{\hsize}{!}{
\includegraphics[viewport= 0 0 340 550,clip]{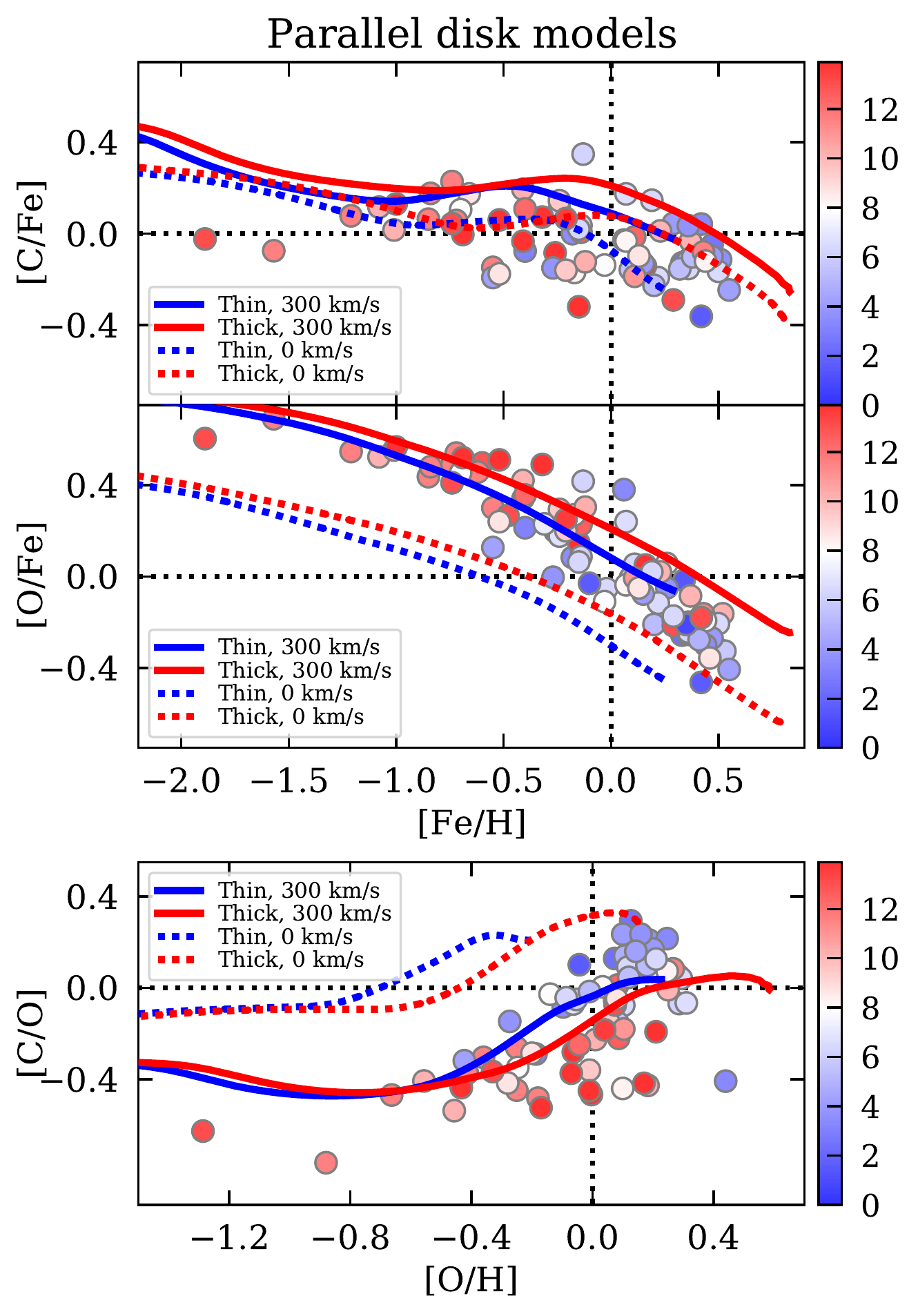}
\includegraphics[viewport= 55 0 340 550,clip]{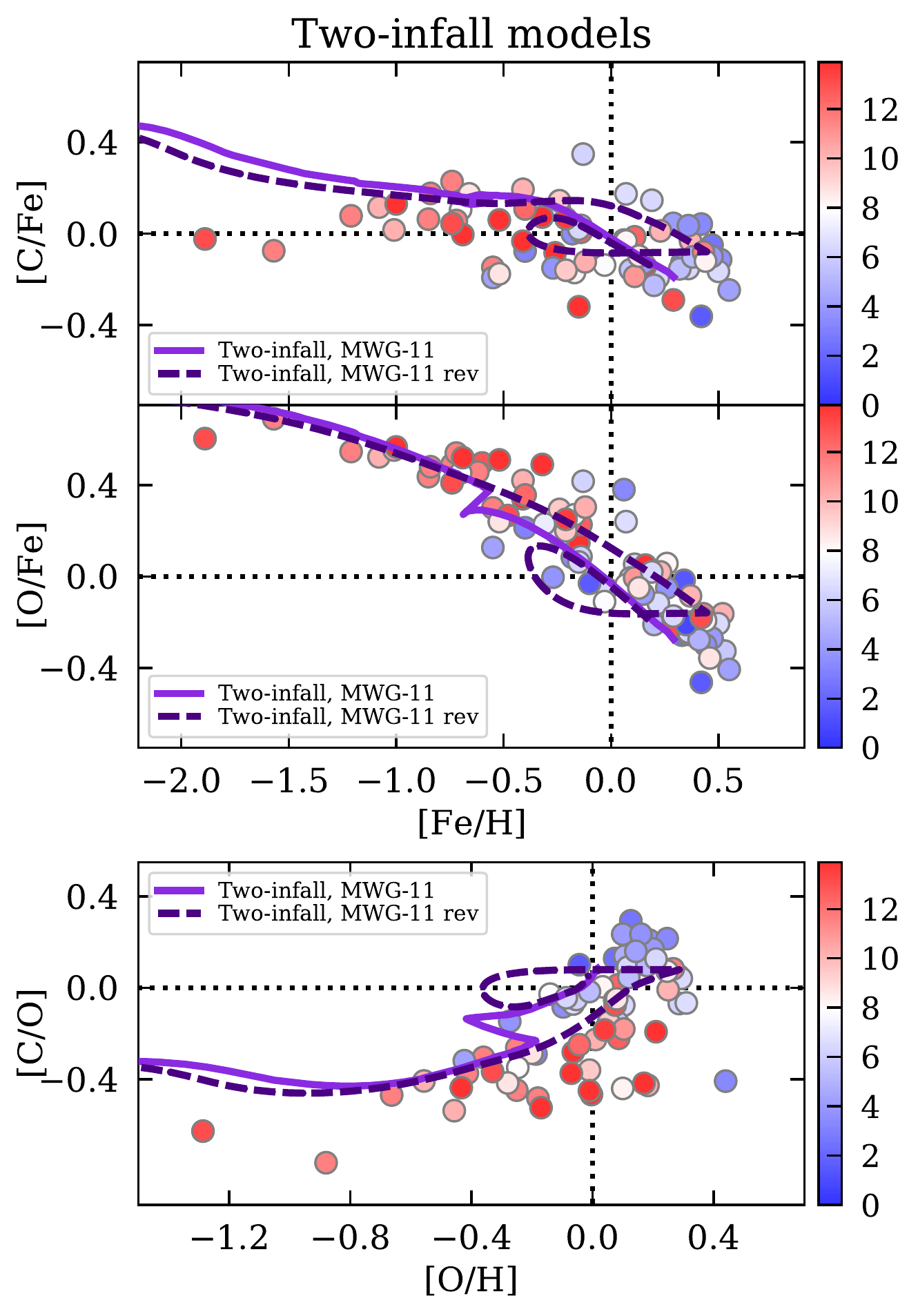}
\includegraphics[viewport= 55 0 380 550,clip]{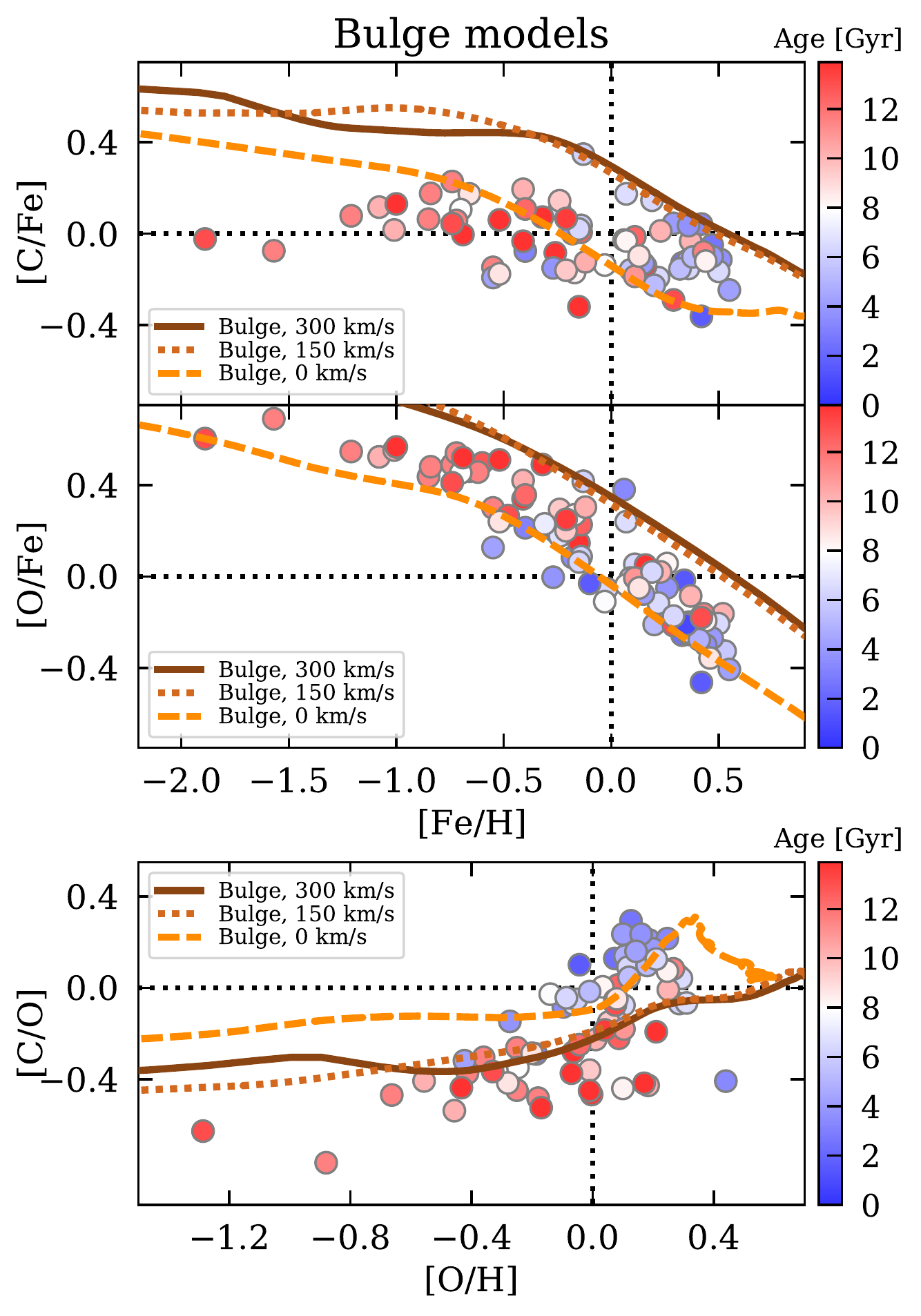}}
\caption{Comparisons of the observed carbon and oxygen abundance trends to the different GCE models from \cite{romano2020}. 
Left-hand side plots: The parallel model for thin disk (blue lines) and thick disk(red lines) by \cite{grisoni2017} with updated stellar nucleosynthetic yields from \cite{romano2019}, here for massive stars rotating at at either zero (dotted lines) or 300\,$\kms$ (solid lines);
Middle plots: The two-infall MWG-11 model from \cite{romano2019} but revised according to the prescriptions in \cite{spitoni2019} (dark purple dashed lines) and without the revision by \cite{spitoni2019} (solid purple lines);
Right-hand side plots:  The bulge models by \cite{matteucci2019} with the nucleosynthetic yields of the MWG-05, MWG-06, and MWG-07 models by \cite{romano2019}, that includes rapidly rotating massive stars at 300, 150, and 0\,$\kms$, respectively. 
\label{fig:models}}
\end{figure*}

\subsection{Comparisons to chemical evolution models}

In the following discussion we will make comparisons of the observed abundance trends to the models of Galactic chemical evolution (GCE) presented by \cite{romano2020} that include yields from low- and intermediate-mass stars from \cite{ventura2013} and yields from rotating high-mass stars from \cite{limongi2018}. The first model is the MWG-11 model of the solar vicinity from \cite{romano2019} that is based on the two-infall models originally developed by \cite{chiappini1997}, but revised (MWG-11 rev) according to the prescriptions of \cite{spitoni2019}. This model includes the stellar nucleosynthetic yields by rotating massive stars that vary with metallicity; at low metallicities  below $\rm [Fe/H]<-1$ the maximum rotational velocity of 300\,$\kms$ was applied, but reduced to 0\,$\kms$ at solar metallicities.  The second model is the parallel thin and thick disk model by \cite{grisoni2017}, in which the two disks evolve independently, but with updated stellar nucleosynthetic yields from massive stars rotating at different rotational velocities (0, 150, or 300\,$\kms$). The last model that we adopt from \cite{romano2020} is the bulge model by \cite{matteucci2019} that assumes that all stars are old and formed in an early starburst, that is, a so-called classical bulge formation scenario (also here with the nucleosynthetic yields of rotating massive stars at different velocities of 0, 150, or 300\,$\kms$). By default the GCE models by \cite{romano2020} were normalised to the \cite{lodders2009} solar abundance scale, but here they are  re-normalised to the \cite{asplund2021} solar abundance scale. 

The different models are shown in Fig.~\ref{fig:models} together with the new abundance results for the microlensed bulge dwarf stars. The upper panel shows the C and O trends with Fe as a reference element, and the lower panel shows the C trends with O as a reference element.

Lastly, before going into the comparison of the GCE models to the observational data, we caution that GCE models, as stated by \cite{romano2020}, by construction, predict the climate, but not the weather, and are only meant to reproduce the average trends and cannot account for the spread in the observed data\footnote{Rumours have it that the core of this wise saying comes from Steve Shore: {\sl "They [GCE models] are a way to study the climate, not the weather, in galaxies."} (as reported by Monica Tosi in some conference proceedings...), (Donatella Romano, private communication).}.

\subsubsection{Parallel thin and thick disk models}

The parallel thin and thick disk models with two sets of yields from rotating massive stars, 0\,$\kms$ and 300\,$\kms$, respectively, are shown in the plots on the left-hand side in Fig.~\ref{fig:models}.
The observed $\rm [C/Fe]-[Fe/H]$ data are relatively well-matched by the model version where the massive stars are non-rotating while the $\rm [O/Fe]-[Fe/H]$ trend is well-matched by the model version in which the massive stars are rotating at 300\,$\kms$. The combined evolution of C and O in the $\rm [C/O]-[O/H]$ plane shows that the best matching models are the ones with yields from massive stars rotating at 300\,$\kms$. This good match is driven by the good match of oxygen to models with yields from rapidly rotating massive stars, and it is difficult to evaluate whether this is also a good match for carbon (probably not). A solution might be that the C yields from rapidly rotating massive stars should be lowered (as the match of oxygen is so good in the $\rm [O/Fe]-[Fe/H]$ plane, and as oxygen is generally believed to have a well-understood nucleosynthetic origin). If that were the case the parallel thin and thick disk GCE models would be able to better match the observed data in the $\rm [C/Fe]-[Fe/H]$ and $\rm [C/O]-[O/H]$ planes. It should be noted that \cite{limongi2018} use the $^{12}$C($\alpha$, $\gamma$)$^{16}$O rate from \cite{kunz2002} which could result in an overproduction of carbon.

\subsubsection{Two-infall models}

The two-infall models shown in the middle column of Fig.~\ref{fig:models} are able to match the observations reasonably well. The best agreement is seen for oxygen in the $\rm [O/Fe]-[Fe/H]$ plane, while they appear to produce slightly too much carbon in the $\rm [C/Fe]-[Fe/H]$ plane (as was also seen for the parallel thin and thick disk models). If the C yields from massive stars could have been reduced, an overall better match in both the $\rm [C/Fe]-[Fe/H]$ and $\rm [C/O]-[O/H]$ could have been reached. It is worth stressing that the models discussed up to now are calibrated on the solar vicinity. In the next section, we compare the data to models specifically tailored on the Galactic bulge.

\subsubsection{Bulge models}

The bulge models shown in the plots in the right-hand side column in Fig.~\ref{fig:models} are not in agreement with the observations. The models with yields from rapidly rotating massive stars are not matching either of the $\rm [C/Fe]-[Fe/H]$ or $\rm [O/Fe]-[Fe/H]$ trends. In both cases the C and O abundances are too elevated in the models. The models with yields from non-rotating massive stars are partly able to follow some parts of the observations, but the overall match is rather poor. The situation gets better when considering the $\rm [C/O]-[O/H]$ plane, but that improvement is clearly due to the fact that the poor matches on the two other planes are cancelling out to some degree. It is clear that the overall appearance of the GCE models are not similar to the observed abundance data in the bulge.    

\begin{figure}
\centering
\resizebox{\hsize}{!}{
\includegraphics{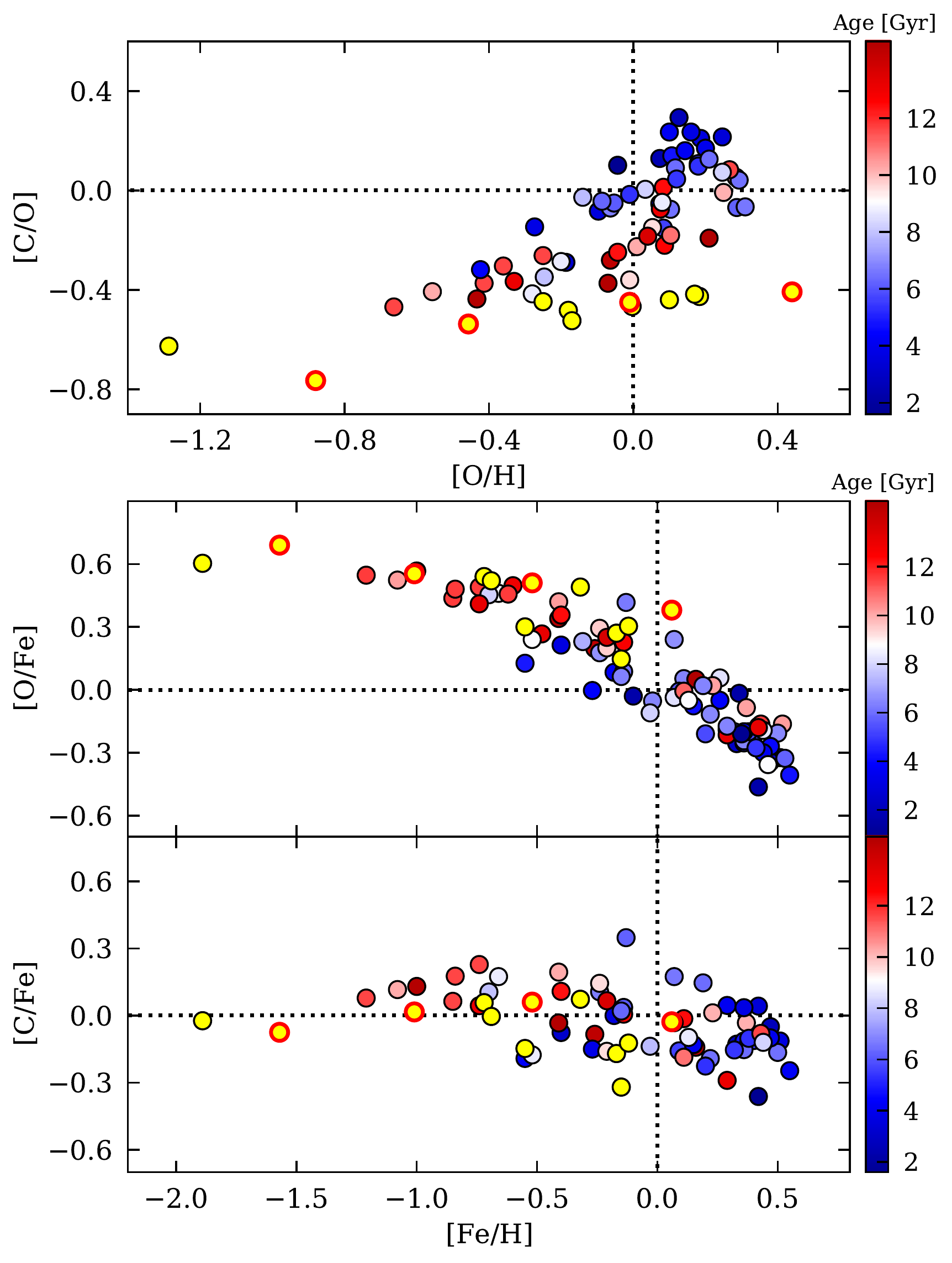}}
\caption{
The $\rm [C/O]-[O/H]$ trend where the low-[C/O] stars are defined and marked by yellow circles.
These are then also marked in the [O/Fe], and [C/Fe] versus [Fe/H] plots.
The four stars that currently have proper motion estimations (given in Table~\ref{tab:propermotion}) are marked with red edges.
\label{fig:ofecfe2}}
\end{figure}

\subsection{The origin of carbon in the bulge}

The overall good agreement between the two-infall and the parallel disk models to the observed abundance trends in the bulge is a striking feature as what we are comparing is bulge abundance trends to GCE models made to simulate the Galactic disk in the solar vicinity. \cite{grisoni2018} expand these  models to other galactocentric radii and even though the star formation rate is much higher at distances closer to the centre of the Galaxy, the overall appearance of the abundance trends does not seem to vary much for the two-infall model.

In the early times of the Milky Way it is clear that the enrichment of the interstellar medium must come from massive stars as it is those that can contribute on short time scales. The good match of the oxygen trends to the models (both parallel and two-infall) with yields from rapidly rotating massive stars  is re-assuring as oxygen is the element that we believe has a very clean and well-understood origin \citep[e.g.][]{woosley1995}. The slight mismatch of the carbon trends would then mean that the prescriptions that went into the yield calculations of massive stars result in too much carbon being produced. From the $\rm [C/O]-[O/H]$ abundance trends we see further that it is when the low- and intermediate mass stars start to contribute to the enrichment that the [C/O] ratio rapidly increases. Alternatively, metallicity-dependent C yields from massive stars can result in higher C production at high metallicities \citep{maeder1992,prantzos1994}.

\cite{romano2020} conclude that at least 60\,\% of the carbon in the disk comes from massive stars, and the fraction is even higher for stars in the bulge.  The agreement between the elemental abundance trends in the bulge and the solar neighbourhood thin and thick disks tells us that not only are the bulge and the disk(s) intimately connected, if not the same, but also that the origin of carbon is the same in both regions. 

It is clear that the solution is not easy, and both observational and theoretical studies come to differing solutions regarding the importance of, and relative contributions of, carbon production from low-, intermediate-, and high-mass stars \citep[e.g.][and references therein]{cescutti2009,mattsson2010,franchini2020}. Our observations seem to favour that the fraction of carbon being made in massive stars is lower, although how much lower we cannot say: more detailed GCE models and stellar yield calculations are needed to explore that.

\subsection{Traces of a classical bulge?}
\label{sec:classicalbulge}

The formation of the bulge has for long been regarded as having occurred on very short time scales \citep[e.g.][]{matteucci1990,ballero2007,cescutti2011,renzini2018}. However, recent observational data suggest that most of the stars in the bulge region could have a secular origin made from disk material \citep[e.g.][]{melendez2008,shen2010}, which is also supported by the results from the microlensed dwarf stars \citep{bensby2017} with their wide range of ages, complex metallicity distribution with two or more peaks \citep[see also, e.g.,][]{hill2011,ness2013,zoccali2017,rojasarriagada2017,rojasarriagada2020}, and detailed elemental abundance trends that mimic the ones of the local thin and thick disks \citep[see also, e.g.,][]{alvesbrito2010,johnson2014,gonzalez2015,jonsson2017}. But that does not mean that we can rule out that a part of the stars in the bulge region do originate from an early intense star formation event that can be associated with a classical bulge scenario. The question is how large that part could be?

In the C and O abundance plots in Figs.~\ref{fig:ofecfe} and \ref{fig:models} there are some stars that appear to deviate from the general abundance trends, maybe even forming a separate trend of their own. In Fig.~\ref{fig:ofecfe2} we plot the abundance trends again but highlight these stars, as well as some stars at lower [O/H] values that align with these deviating stars at higher [O/H], with yellow circles. In the $\rm [O/Fe]-[Fe/H]$ plane these stars tend to be located at slightly more enhanced levels, much in the same way as the bulge in general has been claimed to show slightly more elevated $\alpha$-element abundances than the local thick disk \citep{bensby2013,johnson2014,bensby2017,jonsson2017}. Could it be that these slightly elevated stars actually belong to another subgroup of stars in the bulge region? In the $\rm [C/Fe]-[Fe/H]$ plane these deviating stars are on average less carbon-enhanced than the other stars at similar metallicities. They are more or less perfectly in lock-step with [Fe/H], and maybe at just below solar-metallicity they tend to be under-abundant. Comparing the selected yellow stars in Fig.~\ref{fig:ofecfe2} does not show much agreement to either of the GCE models in Fig.~\ref{fig:models}. A potential match between the GCE bulge models in the plots on the right-hand side in Fig.~\ref{fig:models} and the observations could be had if the C yields from massive stars are decreased significantly.

If the 11 chemically-distinct yellow points in Fig.~\ref{fig:ofecfe2} are drawn from a population with a distinct origin, such as a 'classical bulge', then this might be reflected in distinct kinematics as well. However, a comparison of the galactocentric radial velocities of these stars to the other stars in the sample with ages greater than eight billion years showed no significant differences. Also other parameters such as the age distribution and the distributions in galactic longitude and latitude does not reveal any deviating properties for these stars. So, in conclusion, even though it is tempting to associate these stars with a classical bulge population, more work and possibly also a larger sample of stars is needed to put such connections on firm ground.

\section{Summary}
\label{sec:summary}

We present a detailed analysis of carbon and oxygen for a sample of 91 microlensed dwarf and subgiant stars in the Galactic bulge. Carbon abundances for 70 stars were determined from line synthesis of six \ion{C}{i} in the wavelength range 9060--9120\,{\AA}, and oxygen abundances for 88 stars from line synthesis of the three \ion{O}{i} lines at 7772--7775\,{\AA}. NLTE corrections were accounted for when calculating the synthetic spectra. As the stars have not reached the evolved giant star phase, the estimated C and O surface abundances have not been affected by internal nucleosynthetic burning processes, and therefore reflect the carbon and oxygen abundances of the gas clouds that the stars were born from.  Our main findings are summarised as follows:

\begin{itemize}
\item
The abundance trends that contain oxygen in the Galactic bulge have the same appearance as seen for other $\alpha$-elements. The one distinction is that at super-solar abundances, oxygen continues to decline, whereas other $\alpha$-elements such as Mg, Ca, and Ti, level out and vary in lockstep with Fe towards the highest metallicities. The explanation that has reached highest acceptance is that oxygen has metallicity dependent yields.
\item
Carbon in the Galactic bulge has a similar appearance as seen for Fe, that is, mainly scattering around solar [C/Fe] at all [Fe/H]. When comparing carbon to oxygen, the [C/O]-[O/H] trend is very similar to what is seen for [Fe/H]-[O/H]. 
\item
Both carbon and oxygen trends in the Galactic bulge are very similar to what is observed locally in the solar neighbourhood in the Galactic thin and thick disks. The old part (older than eight billion years) follows the thick disk, while the young part (younger than eight billion years) follow the thin disk.
\item
Galactic chemical evolution models for the thin and thick disks agree well with the observed [C/O]-[O/H] trends in the bulge. Specially tailored chemical evolution models for the bulge, in which the bulge is represented by a single spheroid, do not match the observed data as well. Hence, the observations point to a secular origin of the Galactic bulge, in which the majority of the stellar population in the inner parts of the Galaxy is formed from existing disk populations. Note that this does not rule out the existence of a small classical bulge component. However, we cannot say how large it is.  
\item
Our results do not support the recent suggestions that carbon should to a higher degree originate from massive stars in the bulge than elsewhere in the disk. Instead our results, the agreement between the carbon trends in the bulge and the thin and thick disks, shows that fraction of carbon being made by massive stars in the bulge should be lower, and similar to what is observed in the Galactic disk(s). 
\end{itemize}

\begin{acknowledgement}

T.B. was funded by grant No. 2018-04857 from The Swedish Research Council. J.M. thanks FAPESP (2014/18100-4). We are grateful to Donatella Romano for providing their Galactic chemical evolution models in machine readable format and for providing valuable comments on a draft version of the paper. We also thank the anonymous referee for valuable comments that improved the clarity of the paper. This work has made use of the VALD database, operated at Uppsala University, the Institute of Astronomy RAS in Moscow, and the University of Vienna. This research made use of Astropy, a community-developed core Python package for Astronomy \citep{python_astropy}, Matplotlib \citep{python_matplotlib}, and NumPy \citep{python_numpy}.

\end{acknowledgement}

\bibliographystyle{aa}
\bibliography{referenser}

\end{document}